\documentclass[10pt,journal,compsoc]{IEEEtran}
\ifCLASSOPTIONcompsoc
  \usepackage[nocompress]{cite}
\else
  \usepackage{cite}
\fi
\ifCLASSINFOpdf
  \usepackage[pdftex]{graphicx}
  \graphicspath{{../pdf/}{../jpeg/}}
  \DeclareGraphicsExtensions{.pdf,.jpeg,.png}
\else
\fi
\usepackage{amsmath}
\usepackage{amsmath}
\usepackage{amsfonts}
\usepackage{amssymb}
\usepackage{graphicx}
\usepackage{xcolor, colortbl}
\usepackage{lscape}
\usepackage{array, multirow, multicol}
\usepackage{float}
\usepackage{enumitem}
\begin{document}
%

\title{A Benchmarking Proposal for DevOps Practices on Open Source Software Projects}
%
%
%
%

\author{
José Manuel~Sánchez Ruiz,~\IEEEmembership{Researcher,~Universidad de Sevilla,}

Francisco José~Domínguez Mayo,~\IEEEmembership{Associate Professor,~Universidad de Sevilla,}

Xavier ~Oriol,~\IEEEmembership{Researcher,~Universitat Politècnica de Catalunya,}

José Francisco ~Crespo,~\IEEEmembership{Researcher,~Universitat Politècnica de Catalunya,}

David~Benavides,~\IEEEmembership{Full Professor,~Universidad de Sevilla,}

and Ernest~Teniente,~\IEEEmembership{Full Professor,~Universitat Politècnica de Catalunya}

\IEEEcompsocitemizethanks{\IEEEcompsocthanksitem José Manuel Sánchez Ruiz,  Francisco José Domínguez Mayo and David Benavides are with the Departamento de Lenguajes y Sistemas Informáticos, Escuela Técnica Superior de Ingeniería Informática. Avda. Reina Mercedes s/n. 41012 Seville, Spain.\protect\\
E-mail: \{jsanchez7,fjdominguez,benavides\}@us.es
\IEEEcompsocthanksitem Xavier Oriol, José Francisco Crespo, and Ernest Teniente are with the Departament d'Enginyeria de Serveis i Sistemes d'Informació (ESSI) and inLab FIB, Campus Nord UPC, Jordi Girona 1-3 08034, Barcelona
E-mail: \{jose.francisco.crespo-sanjusto,xavier.oriol,ernest.teniente\}@upc.edu

}

\thanks{Manuscript received April 19, 2005; revised August 26, 2015.}}

%
%

\markboth{Journal of \LaTeX\ Class Files,~Vol.~14, No.~8, August~2015}%
{Shell \MakeLowercase{\textit{et al.}}: Bare Demo of IEEEtran.cls for Computer Society Journals}
%



\IEEEtitleabstractindextext{%
\begin{abstract}
The popularity of open-source software (OSS) projects has grown significantly over the last few years with more organizations relying on them. As these projects become larger, the need for higher quality also increases. DevOps practices have been shown to improve quality and organizational performance. The DORA benchmarking reports provide useful information to compare DevOps practices performance between organizations, but they focus on continuous deployment and delivery to production, while OSS projects focus on the continuous release of code and its impact on third parties. The DORA reports mention the increasing presence of OSS projects as they are widely used in the industry, but they have never been used to measure OSS projects' performance levels. This study reveals that the DORA benchmark cannot be directly applied to OSS projects and proposes benchmarking metrics for DevOps practices, being the first one that adapts the DORA metrics and applies them in the OSS projects context. It analyzes the limitations of the DORA benchmarking and adapts their metrics for new ones that can be applied to OSS projects. The metrics proposed in this study for benchmarking OSS projects include Release Frequency and Lead Time For Released Changes to measure throughput, and Time To Repair Code and Bug Issues Rate to assess stability. In contrast to the DORA reports, where data is collected through manual surveys, in our proposal, data is collected automatically by Performance-Tracker, a tool we developed that retrieves information from public GitHub repositories. This reduces the risk of survey-based data collection. Our study also shows the benchmark feasibility by applying it to four popular OSS projects: Angular, Kubernetes, Tensorflow, and VS Code. The benchmarking results for the OSS projects show that the metric values vary from one project to another show, in general terms, that Angular performs best. In addition, we proposed challenges that address the topics and future works to expand the knowledge and findings of this study. Overall, the findings of the study can help to improve future research on OSS projects and provide a better understanding and challenges of the role of DevOps practices in OSS projects.
\end{abstract}

\begin{IEEEkeywords}
open source software, DevOps, software development, software delivery performance, benchmarking
\end{IEEEkeywords}}

\maketitle

\IEEEdisplaynontitleabstractindextext

%
\IEEEpeerreviewmaketitle

\IEEEraisesectionheading{\section{Introduction}\label{sec:introduction}}

%
%
%
%
\IEEEPARstart{O}{pen}-source software (OSS) projects have grown in popularity over the past years. The number of members that take part in these projects has increased and it is still growing\cite{2023-state-oss-report}. According to a recent study in collaboration with the Open Source Initiative\cite{2023-state-oss-report}, 80\% of organizations increased their use of OSS in the last year, implying a growing reliance on this type of software. The OSS project communities work following indications made by their project administrators in order to develop quality software in a manageable environment. This can sometimes become more difficult as the community grows. 

As OSS projects become larger and more complex, they require higher quality in software development. DevOps \cite{kim2021devops} is an approach that extends agile practices within the collaborative culture to improve the software development and delivery process. Many studies prove that the adoption of DevOps practices provides many benefits such as quality and improved organizational performance \cite{MISHRA2020100308}. Although the core of each definition of DevOps is the interaction between development and  operations, there are different definitions of DevOps \cite{erich2018devops}. However, it also poses challenges for organizations \cite{leite2019survey}. As a result, organizations around the world seek to improve their performance through \textit{benchmarking} and the implementation of DevOps practices \cite{dorareports}. 

Performance is defined in this context by quality and time. It is important to deliver quality products/services to users while producing them fast. The performance of a team, organization or project community can be measured in a time period. The data can change from one period to a different one and it needs to be properly contextualized. The evolution of the performance can be represented using data from different periods.

There are many benchmarking definitions in the literature.  Analyzing various definitions, Anand and Kodali\cite{anand2008benchmarking} describe benchmarking as: "[…] a continuous analysis of strategies, functions, processes, products or services, performances, etc. compared within or between best‐in‐class organizations by obtaining information through appropriate data collection method, with the intention of assessing an organization's current standards and thereby carry out self‐improvement by implementing changes to scale or exceed those standards". According to ASQ\footnote{https://asq.org/quality-resources/benchmarking}, benchmarking is the process of measuring products, services, and processes against those of organizations known to be leaders in one or more aspects of their operations. Therefore, benchmarking provides necessary insights to understand how an organization compares with similar organizations, even if they are in a different business or have a different group of customers. 

In software organizations, a benchmark is a measurable indicator of performance and efficiency, used to evaluate how well a software organization is meeting its goals and objectives and to compare its performance with other software organizations in the same industry.

Since 2011, Google/DORA has published an annual benchmarking report (initially in collaboration with Puppet Labs) on the state of DevOps in the IT industry to understand what practices and capabilities impact on quality and organizational performance in software development. The DORA benchmarking reports propose four different metrics. Two of them, \textit{Deployment Frequency} and \textit{Lead Time For Changes}, are aimed to evaluate throughput of a project, while the other two, \textit{Time To Restore Service} and \textit{Change Failure Rate}, are intended to assess its stability. Intuitively, Deployment Frequency measures how often an organization releases code to production; Lead Time for Changes measures the time it takes for code to go from being committed to running successfully in production; Time to Restore Service is the time it takes to restore service after a service incident or defect that affects end users occurs; and Change Failure Rate is the percentage of released changes that result in a degraded service. The DORA reports provide useful information about DevOps practices in software development from various teams and organizations. However, they are based on surveys that are answered by IT professionals and do not take into account the objectivity found in other methods which automatically extract the information from software deliverables. 

Moreover, the DORA benchmarking is aimed to collect results from projects developed by the IT industry and, thus, it does not provide enough information to assess the impact of DevOps practices in OSS projects. 

OSS projects are different from typical software projects since there is no central operations service for delivering software as a service. Instead, they release code that third parties can instantiate. Thus, they have unique features like:

\begin{enumerate}
\item decentralized and collaborative teams with a distributed design-build process and a lack of explicit project planning and task assignment,
\item reliance on peer review and community production,
\item open source licenses,
\item community-driven development with support from companies in some cases,
\item transparency, and inclusivity, as well as
\item large numbers of free contributors and geographically distributed development \cite{mockus2002two}.
\end{enumerate}

These features differ from those analyzed in the DORA reports. Therefore, the DORA metrics may not be directly applied to benchmark DevOps practices in OSS projects and have to be rethought and adapted to these particular settings.

In addition to the previous claim, it is also worth noting that there is some empirical evidence (such as survey studies, interviews or case studies) that can provide insight into the benefits and challenges of DevOps in software development \cite{erich2017qualitative, caprarelli2020fallacies,diel2016communication}. However, there is still a gap in the evidence about the effectiveness of DevOps principles and practices and the metrics used to measure them. In particular, there is a lack of works that provide quantitative data on the benefits observed from organizations that implement DevOps \cite{https://doi.org/10.1002/smr.1885}.

Our work in this paper proposes benchmarking metrics for DevOps practices on OSS projects, inspired by DORA. We start from an analysis of the limitations of the DORA benchmarking and we propose four metrics that replace the ones of DORA for OSS projects: \textit{Release Frequency} and \textit{Lead Time For Released Changes} to measure throughput; and \textit{Time To Repair} and \textit{Bug Issues Rate} to assess stability. Release Frequency measures how often code is released; Lead Time For Released Changes measures the time it takes for committed code to be released; Time To Repair Code is the time required to fix a defect; and Bug Issues Rate is the percentage of changes requested to correct bugs or repair code. 

We also show the contribution of our benchmark by applying it to four popular OSS projects: Angular
, Kubernetes
, Tensorflow
, and VS Code
. With this, we are able to assess the impact of DevOps in these particular OSS projects. As code release is an activity within the scope of deployment and delivery, the performance values from the DORA report can be used to compare the performance of the evaluated OSS projects with the average performance of organizations in the DORA report.

In contrast to DORA reports, which rely on manual surveys to collect data, our proposal uses a tool we developed called Performance-Tracker to automatically gather data from GitHub public repositories. This is possible thanks to GitHub's open API, which provides access to information on releases, issues, and commits, among others. Our Performance-Tracker tool is based on the code of the Google Cloud Platform Four Keys project\footnote{https://github.com/GoogleCloudPlatform/fourkeys} and can be found in our GitHub repository\footnote{https://github.com/diverso-lab/performance-tracker}. Therefore, data collection in the benchmarking of OSS projects can be automated, while data collection in the DORA reports is manual. 

Summarizing, the main contributions of our paper are the following:
\begin{itemize}
\item \textbf{We show that the DORA benchmark cannot be directly applied to benchmarking DevOps practices in OSS projects}. To achieve it we need to clarify the terminology used, focusing on the continuous release of code and its impact on third parties instead of continuous deployment and delivery to production.
\item \textbf{We propose new metrics to benchmark DevOps practices in OSS projects}. We obtain them by adapting the DORA metrics to the particular issues related to OSS development.
\item \textbf{We show the performance level of four popular OSS projects.} This allows us to compare the performance of these projects and establishes the mechanisms necessary to identify DevOps practices that have an impact on performance.
\item \textbf{We provide a benchmark based on an automatic measurement using project repositories}, thus reducing the risk of survey-based data collection. We provide a new tool named Performance-Tracker that connects to OSS project repositories and applies our performance benchmark for OSS projects.
\item \textbf{We propose a series of challenges} that can be addressed in the future to expand the knowledge that this study provides.
\end{itemize}


Our findings provide insights for future research and offer a better understanding of OSS projects in the context of DevOps. This study is unique as it proposes objective measures of OSS project performance that have not been reported before. All the material and data is available for the sake of open science.

The rest of the paper is structured as follows: Section 2 shows the DORA benchmark metrics limitations when applying to OSS project's particular settings. Section 3 defines benchmark metrics that are tailored to the nature of OSS projects. In section 4, we validate the applicability of the OSS benchmark metrics by conducting a Benchmarking feasibility study. Section 5 addresses the gaps and challenges that we identified in this work. Section 6 provides an overview of research on the performance evaluation of OSS projects and benchmarking methods, such as the State of DevOps Reports. Finally, in Section 6, we offer conclusions and future research directions.

\section{The DORA Benchmark and OSS Projects}
Given the difference in nature between OSS projects and the way software is developed by the organizations in the DORA report, we need to review the terminology used in the last DORA report to accurately describe and adapt each metric to the context of OSS projects. Thus, while the metrics in the "2022 State of DevOps Report" take a broader organizational perspective, considering continuous deployment and delivery code and its impact on services, software development of OSS projects focuses more on the impact of the software released to third parties.

Thus, terms like "deploying code to production" and "running in production" which are used in the DORA report do not apply when deployment is not considered part of the software development process. Instead, we propose measuring \textit{releasing code} for OSS projects. Releasing code refers to making the code changes or additions publicly available in the repository, where they can be used, tested, and potentially incorporated into other projects by third parties. Additionally, the DORA report refers to incidents as "service incidents" or "defects that impact users," while we find it more appropriate to use terms like "bug issue" for OSS projects to refer to the issues that report a bug in the application, since there is no specific service provided or users involved in it.

Another important factor to consider is the environment in which the metrics are measured, which is crucial for evaluating performance and stability. This raises questions about the relevance of differentiating between development (Dev) and operations (Ops) environments and clarifying the concepts of "release", "deploy", and "deliver" when defining metrics for OSS projects. Also, the scope of the OSS projects about the organizations assessed in the DORA reports needs to be defined since the four OSS projects studied in this research do not deploy their code, but only release it to the intended users.

\begin{figure}[h!]
  \includegraphics[width=\columnwidth]{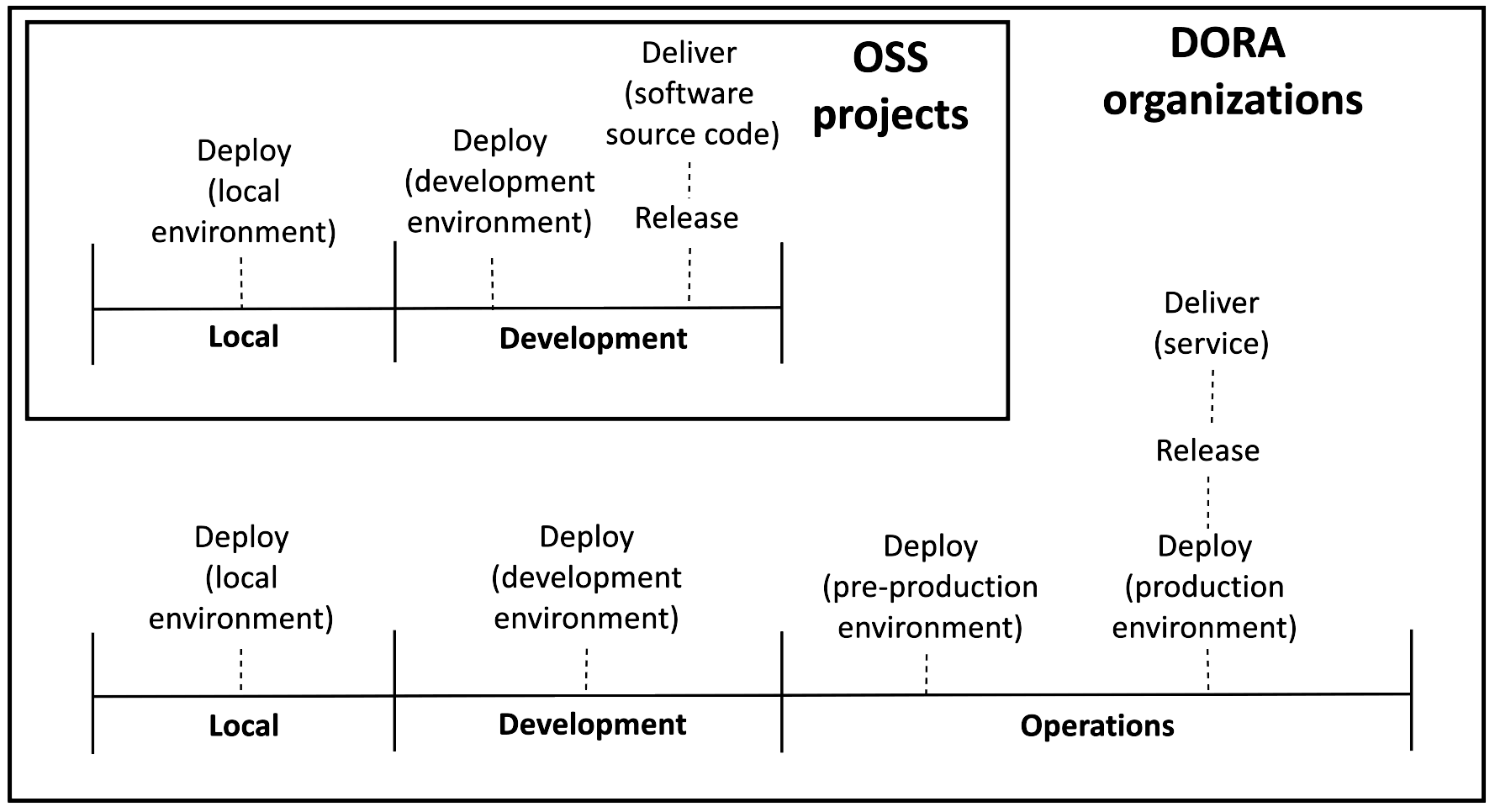}
  \caption{Release, deploy and deliver}
  \label{figure13}
\end{figure}

As shown in Figure \ref{figure13}, OSS projects have different environments which their code goes through, as well as the projects included in the DORA reports. In fact, we can distinguish three different environments:

\begin{itemize}
\item \textbf{Local Environment:} This is where developers write and test code on their own computers. They might deploy the software to test it, but these deployments are not intended for end users, only the developer working on that local environment.
\item \textbf{Development Environment:} This is where software is tested and refined before it is released to a wider audience. The software is usually deployed for testing and a release version is made. In this environment, the release made is considered a delivery in OSS projects, as the software is accessible by end users once the release is made. On the other hand, projects included in DORA reports have one more environment involved in their process.
\item \textbf{Operations Environment:} This is where software is deployed and made available to end users. There are usually two environments involved: pre-production and production. The pre-production environment imitates the production environment with the goal of detecting if new versions of the software work as expected or if they cause any errors. Deployment to the operations environment is usually done by the operations team or a specialized deployment team. Once the software has been deployed to production it is considered that a release has been made and the service is delivered.
\end{itemize}

We can also observe some differences between the terms "deploy", "release" and "deliver".
We define them in order to avoid confusion.

\begin{itemize}
\item \textbf{Deploy:} We can define this term as the action of instantiating a new version of the code, including configuration, testing and installing software on servers.
\item \textbf{Release:} It is considered as the action of making a version of the code available to the users.
\item \textbf{Deliver:} This is the process of offering the software to the users, which includes the process of deploying and releasing the software.
\end{itemize}

In order to explain the differences between the DORA benchmark and the one we developed in this study, we evaluate both benchmark metrics in terms of validity, reliability, accuracy and precision.

\begin{itemize}

\item \textbf{Precision:} This concept refers to the measurements and not to the metric itself. It refers to how wide the measurement thresholds that differentiate one project's performance from another are. In our proposal, in terms of precision, where we consider the same scale as the DORA benchmark, the precision of the OSS projects benchmark would be the same as the DORA benchmark.

\item \textbf{Accuracy:} The accuracy refers to how close the measurement obtained through the metric is to the true value of what we intend to measure. Our OSS project benchmark metrics base their results on the automatic tool Performance-Tracker, in which metrics are defined and it does not rely on interpretations. In contrast, DORA benchmark is based on interpretations that people make of the questions included in the surveys, people might interpret the same question in different ways and give different answers.

\item \textbf{Reliability:} The reliability of a metric refers to whether the results it provides are replicable and consistent. Our OSS project benchmark metrics uses an automatic measuring mechanism (Performance-Tracker tool), which allows us to reproduce the results. These are objective, as they are measured by an automatic tool. In contrast, the DORA benchmark is manual and relies on surveys, where people who participate in surveys can give different answers to the same questions. 

\item \textbf{Validity:} The validity of a metric refers to whether it actually measures what it intends to measure. In other words, it refers to whether the results obtained after the measuring of the metrics are properly interpreted according to what we try to measure. Validity is not a property of the metrics themselves but of the interpretations, we make from the results they provide. In terms of validity, further validation work with experts by both benchmark metrics is needed.

\end{itemize}

In the following, considering the concepts defined above, we go over a more precise and accurate definition of the metrics of the proposed benchmark metrics for OSS projects. In addition, we make sure that the information needed to calculate our metrics can be automatically obtained from GitHub repositories.



\section{Metrics definition}
We created formulas for each of the metrics we propose for benchmarking OSS projects. For each metric, we provide a description, a formula, and the factors we considered in our study, including various measurement methods that we tested. Additionally, for metrics that involve frequency or time, we also include the standard deviation to give more information about the behavior of the OSS project. We explain this further in the relevant metrics that are affected.

We come to our proposal of metrics for DevOps practices on OSS projects, shown in Table \ref{tab:definition_metrics}. The top row reviews the metrics defined in the latest DORA report for software projects led by the industry. The bottom row states our proposal for OSS projects, which is an adaptation of the DORA benchmark to the particularities of the OSS environment. Thus, we propose using \textit{Release Frequency} and \textit{Lead Time For Released Changes} to measure throughput in OSS projects; and \textit{Time To Repair Code} and \textit{Bug Issues Rate} to assess its stability.

\begin{table*}[h!]
\centering
\caption{Definition of Metrics: 2022 State of DevOps Report vs OSS Projects}
\label{tab:definition_metrics}
\begin{tabular}{|p{1.5cm}||p{3.5cm}|p{3.5cm}|p{3.5cm}|p{3.5cm}|} \hline \hline 
\textbf{}   &  \textbf{Deployment Frequency} & \textbf{Lead Time For Changes} & \textbf{Time to Restore Service} & \textbf{Change Failure Rate} \\ \hline 
2022 State of DevOps Report
& For the primary application \textbf{or service} you work on, how often does your organization \textbf{deploy code to production or release} it to end users?  \cite{stateofdevopsreport22}  


& For the primary application \textbf{or service} you work on, what is your lead time for changes (i.e., how long does it take to go from code committed to code successfully \textbf{running in production})? \cite{stateofdevopsreport22} 


& For the primary application \textbf{or service} you work on, how long does it generally take \textbf{to restore service when a service incident or a defect that impacts users occurs} (e.g., unplanned outage or service impairment)? \cite{stateofdevopsreport22}  


& For the primary application \textbf{or service} you work on, what percentage of \textbf{changes to production or released to users result in degraded service} (e.g., lead to service impairment or service outage) and subsequently require remediation (e.g., require a hotfix, rollback, fix forward, patch)? \cite{stateofdevopsreport22}  


\\ \hline \hline

   &  \textbf{Release Frequency} & \textbf{Lead Time for Released Changes} & \textbf{Time to Repair Code} & \textbf{Bug Issues Rate} \\ \hline

OSS projects
& For the primary application you work on, how often does your organization \textbf{releasing code} to end users? 





& For the primary application you work on, what is your lead time for changes (i.e., how long does it take to go from code committed to code successfully \textbf{released in the project repository})?





&  For the primary application you work on, how long does it generally take \textbf{to repair the code when an bug or a defect that impacts users occurs} (e.g., bug issues)?





&  For the primary application you work on, what percentage of \textbf{changes requested imply a bug correction or code repair} (e.g., malfunction of the application)?





\\ \hline
\end{tabular}
\end{table*}

It is important to remark that to measure the performance level of a project using our proposal, one has to establish a period in which the metrics will be measured.




\subsection{Release Frequency}
Deployment Frequency is one of the DORA throughput metrics. It measures how often an organization deploys code to production or releases it to end users.

Given this description of the metric and its differences from the Deployment Frequency metric defined by DORA, we can apply the following formula (\ref{formula22}) to measure the Release Frequency of the projects.

\begin{equation}\label{formula22}
f_{r} = \cfrac{\sideset{}{_{i=0}^{n_{r}}}\sum
t_{r_{i+1}}-t_{r_{i}}}{N}
\end{equation}

Being \(t_{r_{i}}\), \(t_{r_{i+1}}\) the date of two consecutive releases,
 \(n_{r}\) the number of releases made in the period,
 \(N = n_{r}-1\) the number of releases (minus 1) and being \(f_{r}\) the release frequency. Dates can be measured as timestamps to facilitate date operations.

Formula (\ref{formula22}) calculates the average time between consecutive releases during a specific period. The same result can be achieved by dividing the time difference between the release date of the first and last releases by the number of releases minus one. However, we use Formula (\ref{formula22}) because it allows us to obtain a set of data that includes the time between every two consecutive releases. This data enables us to measure the standard deviation of the time between releases, which provides a measure of how much the time between releases varies during the period.

The standard deviation reflects the regularity of the releases made. In these lines, a high standard deviation indicates that the releases are not made regularly, so the number of days between two consecutive releases might vary noticeably. On the contrary, a low standard deviation indicates a more regular Release Frequency (e.g. a release made every week). In this case, a time-based release strategy \cite{timebasedreleases} would have a standard deviation close to zero, as the time between every release would be the same. We define the formula as follow:

\begin{equation}\label{formula3}
d_{fr} = \sqrt{\cfrac{\sideset{}{_{i=0}^{N}}\sum
t_{rd_{i}}-m_{rd}}{N}}
\end{equation}

Being \(t_{rd_{i}}\) the time between two consecutive releases (calculated before for the Release Frequency as \(t_{ri+1} - t_{ri}\)),
 \(m_{rd} = f_{r}\) the mean of times between consecutive releases (which equals the Release Frequency),
 \(N = n_{r}-1\) the number of releases (minus 1) and being \(d_{fr}\) the standard deviation of the release frequency.

As regards the Release Frequency formula we define in this paper, we found two exceptions in which the release frequency can not be calculated:

\begin{itemize}
\item \textbf{There is only one release in the period:} We can not measure how long it takes to make a release after the first one because there are no more releases. In this case, an approximation to the result would be to estimate that the Release Frequency is equal to the period.
\item \textbf{There are no releases in the period:} We can not calculate the release frequency because there are no releases at all in the period. This case only allows us to confirm that the Release Frequency is longer than the period.
\end{itemize}

\begin{figure}[h!]
  \includegraphics[width=\columnwidth]{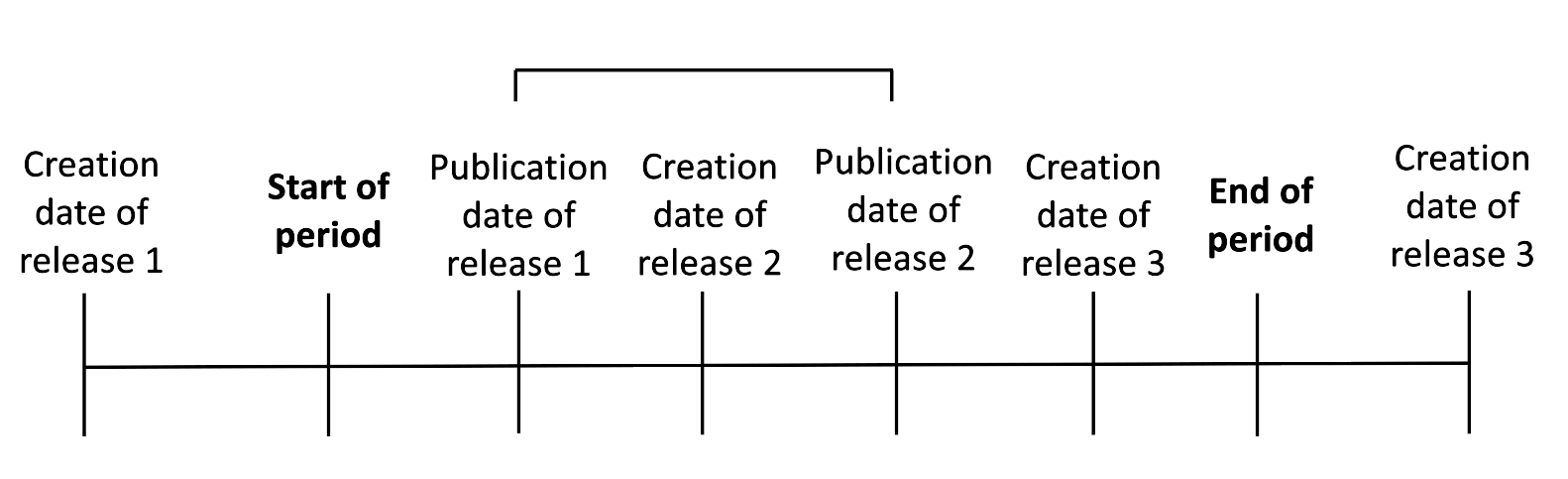}
  \caption{Release Frequency example}
  \label{figure1}
\end{figure}

In this study, we consider that a release is made whenever a GitHub release is published in the given repository. We must take into account that releases have a creation date and a publication date. The releases included in the measurement are the ones that where published in the period. Figure \ref{figure1} shows an example scenario. In this example, releases are consider as follows:

\begin{itemize}
\item \textbf{Release 1} would be included as it was published in the period (it was created outside the period).
\item \textbf{Release 2} would be included as it was published in the period (it was also created in the period). 
\item \textbf{Release 3} would not be included as it was published outside the period (it was created in the period). 
\end{itemize}
  
\subsection{Lead Time For Released Changes}
\begin{figure}[h!]
  \includegraphics[width=\columnwidth]{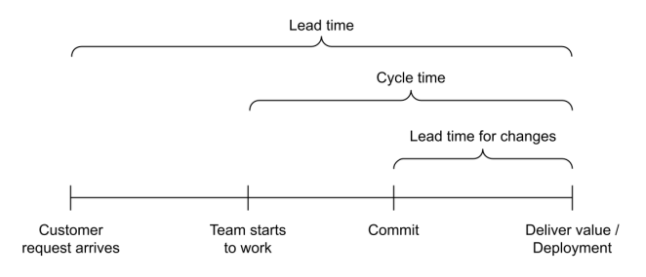}
  \caption{Lead Time, Cycle Time and Lead Time For Changes}
  \label{wpcust}
\end{figure}

Lead Time for Changes is a DORA metric for measuring throughput. For this metric, there are three different concepts that refer to three different metrics whose names can lead to confusion. For this reason, in the following, we include a description of these concepts and the differences between them. Figure \ref{wpcust} represents the relation between these three metrics:

\begin{itemize}
\item \textbf{Lead Time} measures the average time between a change is requested and it is deployed in production. 
\item \textbf{Cycle Time} measures the average time between the moment in which the team starts to work in a change and it is deployed in production. 
\item \textbf{Lead Time For Changes} measures the average time it takes to go from code
committed to code successfully running in production. 
\end{itemize}

 Therefore, these three metrics represent three different time measures of the development and operations process. The DORA project includes the Lead Time For Changes as one of its metrics. 

In this metric, we measure the time it takes to go from committed code to code successfully released, naming this metric Lead Time For Released Changes. To calculate this metric, we measure the time between the last commit for an issue made and the deployment including the issue made. We represent this formula as follows:
\begin{equation}\label{formula5}
l_{t} = \cfrac{\sideset{}{_{i=0}^{n_{s}}}\sum
t_{r}-t_{c_{i}}}{n_{s}}
\end{equation}

Being \(t_{r}\) the date in which the change is released (version of the code that includes the change is released), \(t_{c_{i}}\) the date in which the last commit for an issue is made,
\(n_{s}\) the number of changes requested in a given period that lead to changes in the code and \(l_{r}\) the lead time for released changes.

In the case of not having any changes made to the code during the measuring period, this metric could not be calculated because of lack of data. 

In this metric, the standard deviation represents the differences in time between committing and releasing code from different issues. A high standard deviation reflects that the time it takes from going to committed code to release code for each issue raised might vary a lot, taking some issues more time than others. A low standard deviation indicates that the time it takes from going to commit code to released code for each issue is similar and does not vary much. We propose the following formula:

\begin{equation}\label{formula6}
d_{l} = \sqrt{\cfrac{\sideset{}{_{i=0}^{n_{s}}}\sum
t_{ld_{i}}-m_{ld}}{n_{s}}}
\end{equation}

Being \(t_{ld_{i}}\) the time it takes to go from committed code to released code for each change committed (calculated before for the Lead Time For Released Changes as \(t_{r} - t_{c_{i}}\)),
 \(m_{ld} = l\) the mean of the time it takes to go from committed code to released code (which equals the Lead Time For Released Changes),
 \(n_{s}\) the number of changes requested in a given period that
lead to changes in the code and being \(d_{l}\) the standard deviation of the lead time for released changes.

It can become difficult to measure Lead Time For Released Changes metric, as every OSS project and its communities work in different ways, so trying to generalize the way we measure the metric can lead to imprecise results that are not true to reality. In order to approach the metric in an appropriate way, we studied different possibilities. As explained above, the Lead Time For Released Changes should represent the average time it takes to go from committed code to code released.

\begin{figure}[h!]
  \includegraphics[width=\columnwidth]{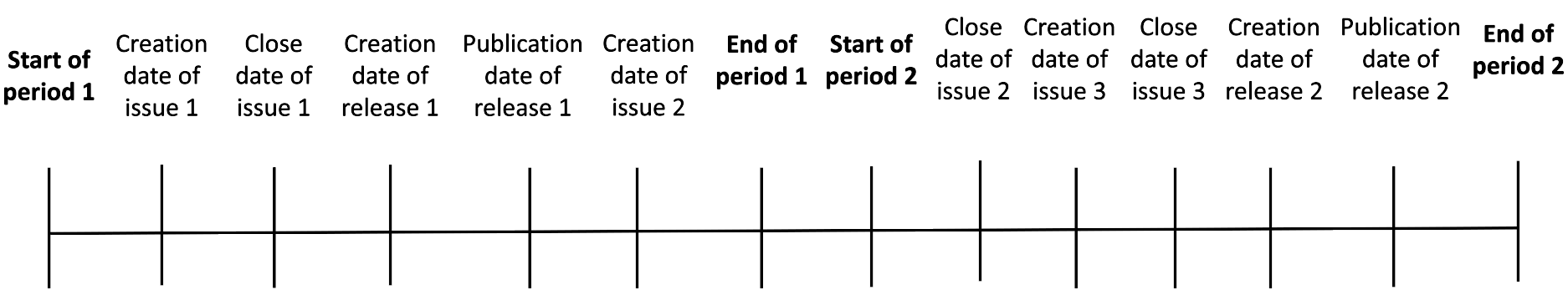}
  \caption{Lead Time For Released Changes example}
  \label{figure5}
\end{figure}

One of the main problems for this metric is knowing what issues to include in the period that is being measured. We need to define which issues belong to a period. At first, we only included issues that were opened and closed within the period. That leads to results in which the Lead Time For Released Changes was never longer than the period. It was not an accurate metric as we were leaving out of it the issues created in other periods but closed in the actual one or created in this period but closed in the future (the changes made for an issue can be already committed but the issue is still open, it can be closed in the future). Figure \ref{figure5} shows an example scenario. In this example, releases are consider as follows:

\begin{itemize}
\item \textbf{Issue 1} would be included in period 1 (it was created and closed in that period). 
\item \textbf{Issue 2} would not be included in any period (it was created in period 1 but closed in period 2).  
\item \textbf{Issue 3} would be included in period 2 (it was created and closed in that period). 
\end{itemize}

\begin{figure}[h!]
  \includegraphics[width=\columnwidth]{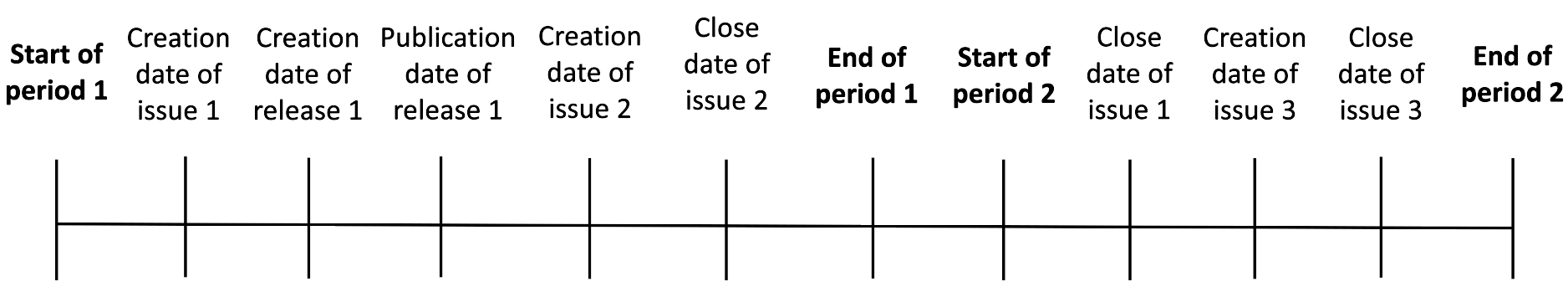}
  \caption{Lead Time For Released Changes example 2}
  \label{figure6}
\end{figure}

Another option would be to take only the issues opened in the period, but some of those issues might not be released yet, so there is no way to measure the Lead Time For Released Changes with those issues. In addition, it would not represent reality in a proper way: the commits and releases of those changes could be made in the future, not properly representing the Lead Time For Released Changes of the actual period that is being measured (Figure \ref{figure6}).

\begin{itemize}
\item \textbf{Issue 1} would be included in period 1 (it was created in period 1 and closed in period 2). 
\item \textbf{Issue 2} would be included in period 1 (it was created and closed in period 1). 
\item \textbf{Issue 3} would be included in period 2 (it was created and closed in that period but it was not deployed yet).
\end{itemize}

It became more obvious that the focus should be on the time in which the last commit is made and the time in which the release that includes the changes is made. We decided to include all the issues that were released in the period. In order to do that, we can take all of the releases published in the period (whether they were created in the period or not) and include every issue whose code was included in that release. It is difficult to know exactly which issues were included in a certain release, so we decided to include all of the issues whose last commit made in the period was made between the creation of that release and the previous one created, as an approximation. This is the final approximation used for this metric. The following diagrams show some examples (Figures \ref{figure7}, \ref{figure8}, \ref{figure9} and \ref{figure10}): 

\begin{itemize}
\item \textbf{Issue 1} would be included in the period (its last commit was made before the creation of release 1, included in the period).  
\item \textbf{Issue 2} would be included in the period (using its last commit made before the creation of release 1, the only release included in the period).

\begin{figure}[H]
\includegraphics[width=\columnwidth]{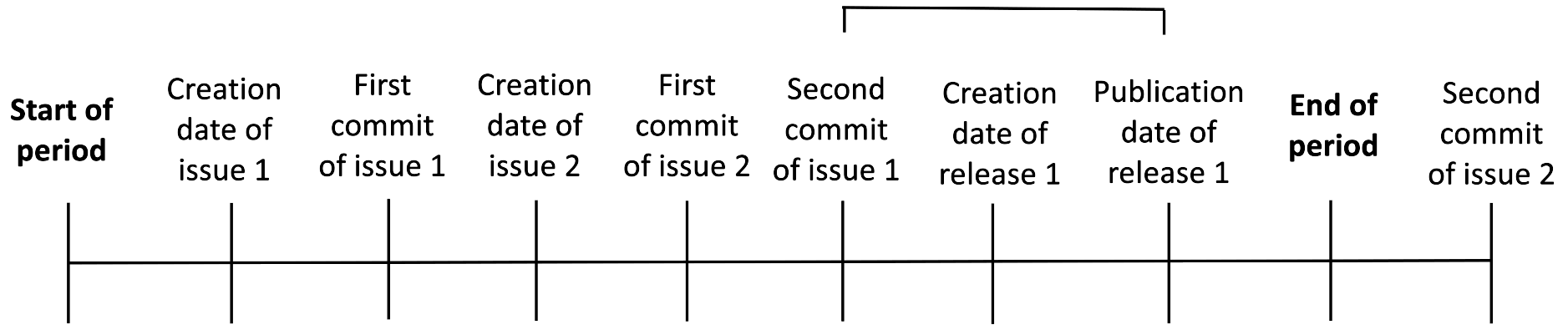}
  \caption{Lead Time For Released Changes example 3}
  \label{figure7}
\end{figure}

\item \textbf{Issue 1} would be included in the period (although its first commit was made before the creation of release 1, the last commit was made before the creation of release 2, so it is considered to be included in that release).  
\begin{figure}[h!]
  \includegraphics[width=\columnwidth]{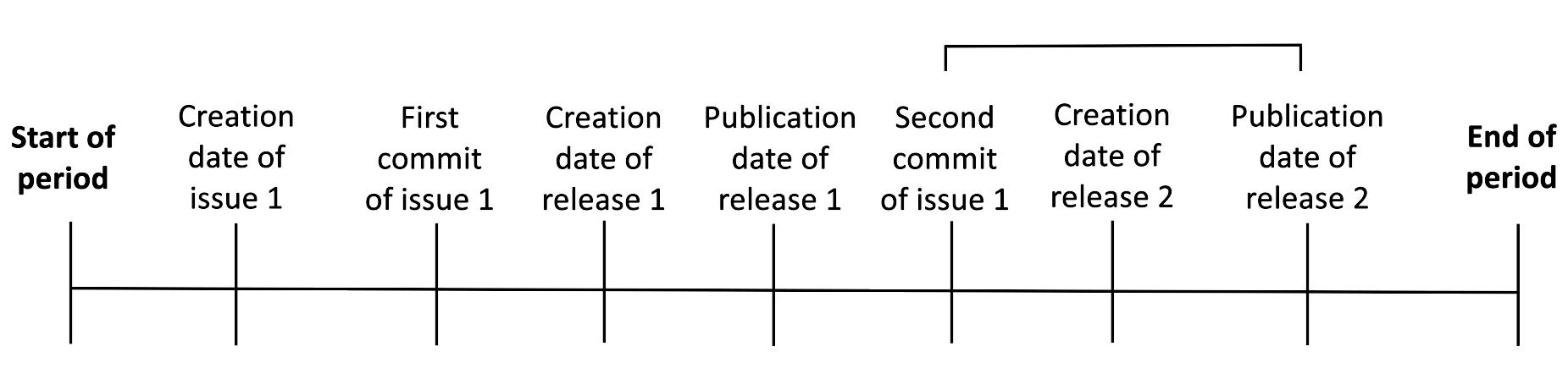}
  \caption{Lead Time For Released Changes example 4}
  \label{figure8}
\end{figure}

\item \textbf{Issue 1} would be included in the period as part of the code included in release 1 (its last commit was made before the creation of release 1, included in the period, although it was published after release 2). 
\item \textbf{Issue 2} would be included in the period as part of the code included in release 2 (its last commit was made after the creation of release 2). 
\begin{figure}[h!]
  \includegraphics[width=\columnwidth]{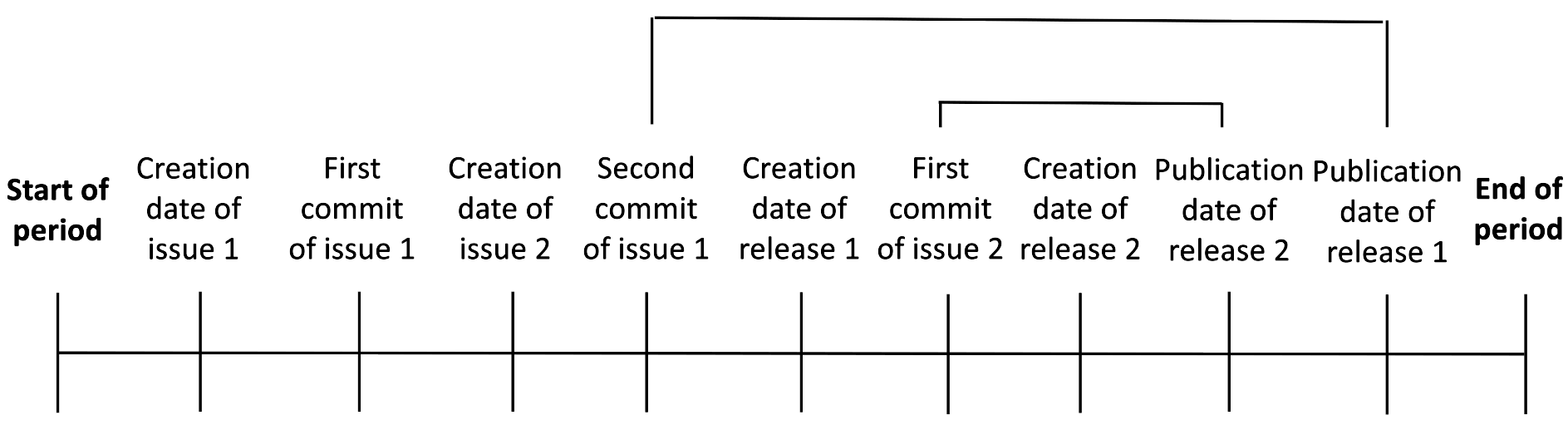}
  \caption{Lead Time For Released Changes example 5}
  \label{figure9}
\end{figure}

\item \textbf{Issue 1} would be included in the period as part of the code included in release 1 (its last commit was made before the creation of release 1, although it was made outside the period). 
\item \textbf{Issue 2} would not be included in the period as its only commit was made after the creation of release 1 (the only release included in the period).
\begin{figure}[h!]
  \includegraphics[width=\columnwidth]{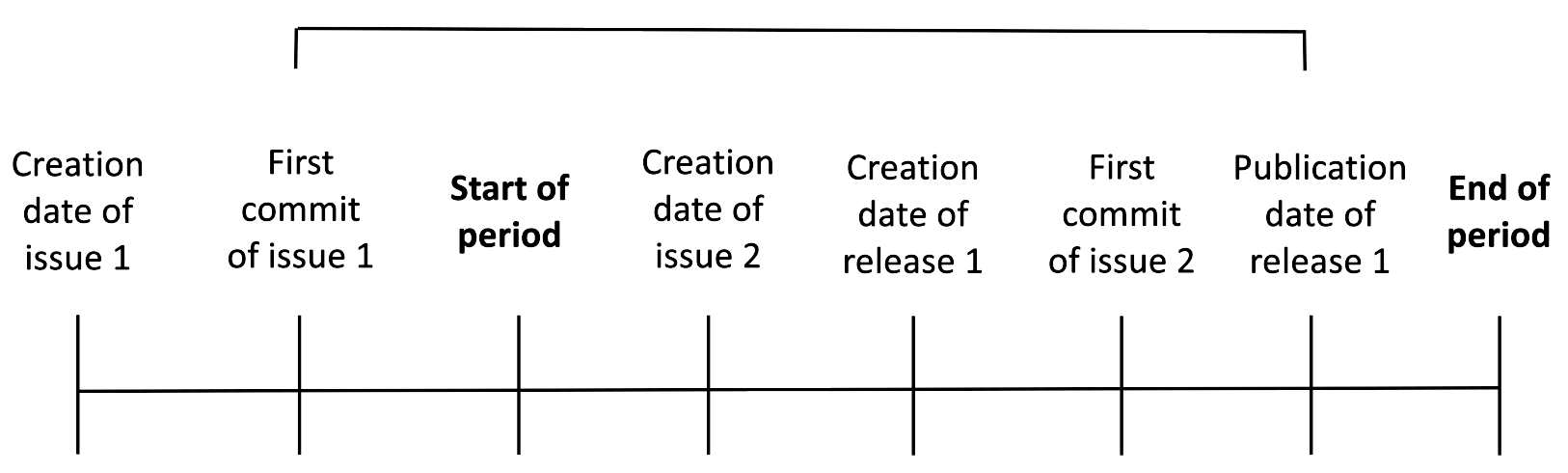}
  \caption{Lead Time For Released Changes example 6}
  \label{figure10}
\end{figure}

\end{itemize}

\subsection{Time To Repair Code}
In the context of the DORA reports, Time To Restore Service is a metric used to measure stability. It measures how long it typically takes to restore service after an bug or defect that impacts users.

However, in OSS projects, this metric is not directly applicable as there is no central operations service where the software is deployed and the team is responsible for maintaining the quality of the service. Instead, OSS projects release versions of the software that can be used by third parties. Therefore, the Time To Restore Service metric in OSS projects is adapted to the Time To Repair Code, which refers to the time it takes to release a new version of the code that includes changes that correct bugs.

In order to measure this metric we tried to develop an approximation based on the information we could retrieve from GitHub. The time it takes to repair a bug would be the time passed between the moment in which the bug issue is raised and the moment in which the changes that solve that bug are released in a new version of the code. We define the following formula:

\begin{equation}\label{formula7}
m_{r} = \cfrac{\sideset{}{_{i=0}^{n_{e}}}\sum
t_{r}-t_{e_{i}}}{n_{e}}
\end{equation}

Being \(t_{r}\) the moment in which the changes that solve the bug are released, \(t_{e_{i}}\) the moment in which the team is notified about the bug (the issue reporting the bug is opened), \(n_{e}\) the number of bugs that users reported in a period of time and lead to changes in the code and \(m_{r}\) the time to repair code. 

In the case of not having any changes that solve bugs made to the code during the measuring period, this metric could not be calculated because of lack of data. 

In this metric, the standard deviation reflects the variations in the time it takes to solve bugs. A high standard deviation represents large variations, meaning that the time it takes to solve different bugs varies. A low standard deviation represents a little variation in the time it takes to solve different bugs. We propose the following formula:

\begin{equation}\label{formula8}
d_{m} = \sqrt{\cfrac{\sideset{}{_{i=0}^{n_{e}}}\sum
t_{md_{i}}-m_{md}}{n_{e}}}
\end{equation}

Being \(t_{md_{i}}\) the time it takes to solve a bug (calculated before for the Time To Repair Code as \(t_{r} - t_{e_{i}}\)),
 \(m_{md} = m\) the mean of the time it takes to solve bugs (which equals the Time To Repair Code),
 \(n_{e}\) the number of changes introduced in the code that were meant to solve a bug and being \(d_{m}\) the standard deviation of the time to repair code.

\begin{figure}[h!]
  \includegraphics[width=\columnwidth]{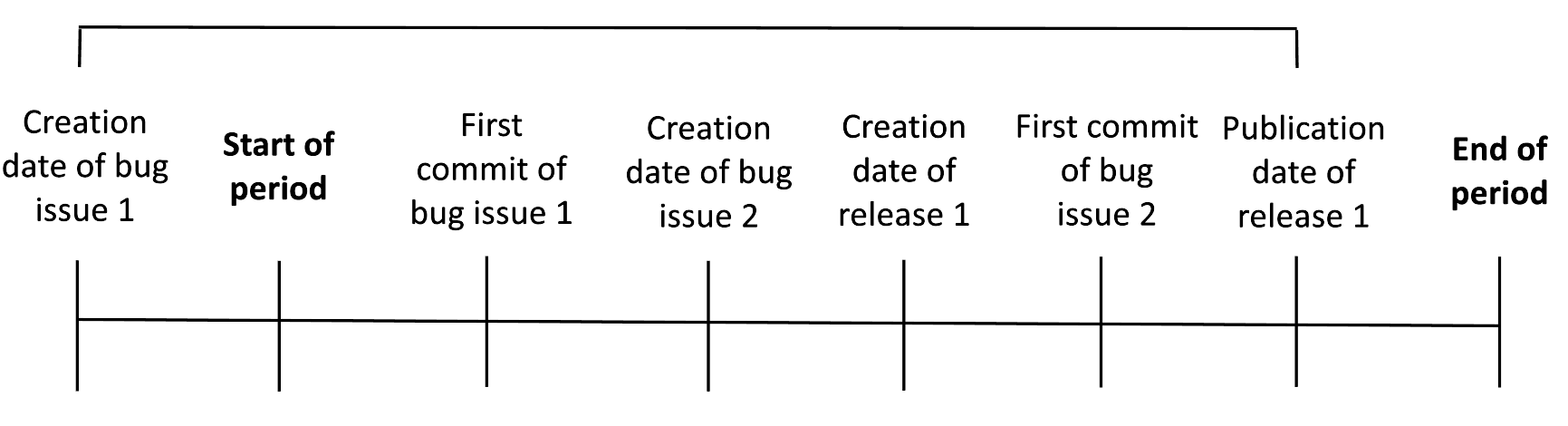}
  \caption{Time To Repair Code example}
  \label{figure11}
\end{figure}

We can notice that Time To Repair Code metric is similar to Lead Time For Released Changes. It is actually a version of that metric in which the only issues included are those which refer to bugs and the time measured is between the issue is created and the changes that correct the bug are released. Figure \ref{figure11} shows an example scenario. In this example, bug issues are consider as follows:

\begin{itemize}
\item \textbf{Bug issue 1}  would be included in the period as part of the code included in release 1 (its last commit was made before the creation of release 1, although it was made outside the period).
\item \textbf{Bug issue 2}  would not be included in the period as its only commit was made after the creation of release 1 (the only release included in the period). 
\end{itemize}

\subsection{Bug Issues Rate}
The Change Failure Rate is a metric used to measure stability in the DORA reports. It determines the percentage of changes made to production or released to users that result in degraded service. Similar to the Time To Restore Service metric, it can be challenging to identify when an bug occurs and which changes caused the bug when analyzing OSS projects that deliver a product.

Again, we developed an approximation named Bug Issues Rate, in which we measure the ratio that represents the number of bug issues raised in a period compared to the total number of issues raised. We define the following formula:

\begin{equation} \label{formula9}
q = \cfrac{n_{e}}{n_{s}}
\end{equation}

Being \(n_{s}\) the number of issues opened in a period of time, \(n_{e}\) the number of bug issues that happened in a period of time and \(q\) the bug issues rate. This metric represents the relation between the number of issues opened in a period and the percentage of them that are bug issues (are labeled as a bug).
 
 
 


\section{Benchmarking Feasibility Study}
For our study to be carefully designed and implemented to provide accurate, unbiased and informative results, we applied the recommendations and guidelines of research studies on benchmarking methods \cite{weber2019essential}. Benchmarking can be conducted (1) by the authors of new methods to demonstrate performance improvements or other benefits; (2) by independent groups interested in systematically comparing existing methods; (3) or organized as "Neutral"’ benchmarking studies \cite{boulesteix2013plea, boulesteix2017necessity}. In our case, we performed a benchmarking study without any perceived bias, focusing on the comparison itself, and comparing methods that are applied in several OSS projects. In addition,  We followed reproducible research best practices to make the procedure performed and the data publicly available. The study was organized in the following phases: 

\begin{itemize}
\item \textbf{Planning the benchmarking}: In this phase, we clearly defined the purpose and scope of the benchmark, developed appropriate measurement methods, and identified representative datasets from OSS projects. We also chose relevant parameter values and software versions, which guided the design and implementation of the study. Specifically

\begin{enumerate}
[leftmargin=0.25cm]
\item We \textbf{developed a measurement instrument} to capture key metrics for each of the OSS projects in the sample, as described in Section 3
\item We defined criteria for \textbf{selecting OSS project repositories}, and wrote a draft description of each practice and tool in order to provide an accurate description of their practices. This activity was carried out in parallel with the definition of the metrics, and the draft was checked by all authors of this paper.
\end{enumerate}

\item \textbf{Running and reporting the benchmarking:} In this phase, we assessed the performance of OSS projects using the quantitative metrics we proposed. We analyzed the practices and tools used in these projects to gain a deeper understanding of their performance. To interpret the results, we used DORA benchmarking performance levels as a reference point. This helped us classify the OSS projects as high, medium, or low performers and provided context for our recommendations. Overall, we provided recommendations based on a community perspective of OSS projects.
\end{itemize}
Finally, we provided a summary of the most relevant results and discussed the main highlights of potential threats to the validity of our study, as well as its limitations.

\subsection{Planning the benchmarking}
\subsubsection{Developing a measurement instrument}
We developed a software tool named Performance-Tracker based on the Google Cloud Platform Four Keys project\footnote{https://github.com/GoogleCloudPlatform/fourkeys}. Our tool measures the four metrics for OSS projects, which are then calculated based on the formulas we defined in this paper.
Existing software measurement tools can gather data from repositories automatically such as Google Cloud Platform Four Keys and Apache DevLake, but they must be integrated with the software organization's infrastructure, including GIT workflows, issue management systems, code repositories, and CI/CD automation tools.


Then, the variability in GIT workflows poses a challenge in collecting data \cite{10.1016/j.infsof.2021.106811}, as the data capture depends on the particular GIT workflow. 

\subsubsection{Selection of the OSS Projects to Benchmark} 

In our study, we selected four popular OSS projects. Our goal was to measure and compare them. To be included in our study, we defined specific criteria: the projects had to have over 1,500 contributors, be created before 2016, have more than 20,000 issues, and have over 10,000 pull requests. We gathered data from GitHub, a public software development platform, during a six-month period from July 1st, 2022 to December 1st, 2022, using the GitHub API to collect information such as releases, commits, and other data from the projects within that time period. We chose a six-month duration for data collection, as it provided sufficient time to obtain the necessary data for accurately representing each project's results. We refrained from using extended measurement periods to prevent an excessive number of requests, given that the GitHub API imposes a 5000 requests per hour limitation for each user.

According to the previous criteria, the selected OSS projects are Angular\footnote{https://github.com/angular/angular}, Kubernetes\footnote{https://github.com/kubernetes/kubernetes}, Tensorflow\footnote{https://github.com/tensorflow/tensorflow} and VS Code\footnote{https://github.com/microsoft/vscode}. We conducted analysis and research on each project to better understand how they work. A description of each project and relevant data appear in Table \ref{tab:openprojectsinformation}.

\begin{table*}[h!]
\centering
\caption{OSS projects information as for December 2022}
\label{tab:openprojectsinformation}
\begin{tabular}{|p{2cm}||p{2cm}|p{2cm}|p{2cm}|p{2cm}|} \hline 
\textbf{Project} & \textbf{Contributors} & \textbf{Created} & \textbf{Issues} & \textbf{Pull Requests} \\ \hline \hline 
Angular & 1,573 & 2014-09-18 & 24,017 & 21,467 \\ \hline
Kubernetes & 3,195 & 2014-06-06 & 41,028 & 69,494 \\ \hline
Tensorflow & 3,127 & 2015-11-07 & 34,910 & 21,189 \\ \hline
VS Code & 1,655 & 2015-03-09 & 135,667 & 12,899\\ \hline
\end{tabular}
\end{table*}

\begin{itemize}
\item \textbf{Angular:} Angular is an OSS framework used worldwide for the development of front-end and single-page applications. Written in Typescript and supported by Google, it was first released in 2016.
In the Angular repository, it is possible to know when a new version of the code has been released. These are tags that define different versions of the code. Given the workflow used in the Angular repository, new versions of the code are released including the changes made since the last version was released. 
When a pull request is made, the changes are checked by an authorized user and a merge is made, introducing the changes into the Angular main branch. 
We take the issues considered bug issues as representations of the moments in which a bug in the system was detected. These can be described as those issues that refer to an bug (usually labeled as a bug). 
Tests are usually included in every change made to the code to check that everything works as expected and it is, in many cases, required when contributing to an OSS project. This reduces the risk of committing code that produces errors in the application. 
\item \textbf{Kubernetes:} Created in 2014 by Google, Kubernetes (commonly referred to as K8s) is a platform to automatize the administration of services. It allows users to automatically deploy systems and manage containerized applications, among other uses. 
Releases of the code are made automatically by a bot in GitHub. The releases include the version of the code in the format vXX.YY.ZZ. 
Issues are created by the community and usually triaged by a bot, that assigns the needed tags that help to classify the issue. Many commits are made linked to pull requests but not to the issues, which makes it difficult to know the number of commits made for a certain issue without accessing pull requests. For bug issues, we can look for the kind/bug label that identifies them. 
\item \textbf{TensorFlow:} TensorFlow is an OSS library for dataflow and differentiable programming across a range of tasks. It is a symbolic math library and is also used for machine learning applications such as neural networks. It was developed by the Google Brain team and was first released in 2015. Now it is maintained by the TensorFlow organization.
Releases in this repository work as in the previous ones. However, we can point out that the releases are made with less frequency, but there are multiple releases made in the same day for different versions of the code, which can lead to confusion in the metrics. 
When a pull request is made, the changes are checked by an authorized user and a merge is made, introducing the changes into Tensorflow main branch. For bug issues, we can look for the type:bug label that identifies them. 
\item \textbf{VS Code:} VS Code is an IDE developed by Microsoft for Windows, Linux, macOS and web. It has been in continuous development since it first release in 2015. The editor includes code depuration and Git integration among other features. 
Similar to the previous projects, VS Code uses GitHub releases in its repository. However, it does not use pre-release tag, as other OSS projects do. Because of that, every release is considered a release to production. 
When a pull request is made, the changes are checked by an authorized user and a merge is made, introducing the changes into the VS Code main branch. 
Bug issues in the VS Code repository are labelled as bug.
\end{itemize}

\subsection{Running and reporting the benchmarking}
We evaluated the performance of open-source software (OSS) projects using the metrics we proposed. We also examined the practices and tools employed in these projects to obtain a better understanding of their performance.

To reproduce the results of our study, we have provided the software tool we used on this project in the following repository\footnote{https://github.com/diverso-lab/performance-tracker}. This software was created and utilized for this research, and it was adapted to support OSS benchmarking metrics from the Google Cloud Platform Four Keys project's code. To collect data and compute the metrics we just have to download the software and edit the .env file, which includes some variables, as described in the repository.

To interpret the results, we used DORA benchmarking performance levels as a reference point. Table \ref{tab:performancelevels} shows how the DORA reports classify the performance levels of the projects according to metric results. This helped us classify the OSS projects as high, medium, or low performers and provided context for our recommendations. Table \ref{tab:performanceresults} includes, for each project, the name of the project and the data obtained after measuring each metric, including the standard deviation of the metrics Release Frequency, Lead Time For Released Changes and Time To Repair Code of the four OSS projects included in this study. The metric Bug Issues Rate does not have standard deviation as it is a ratio and not a mean calculated from a set of values. The standard deviation represents how the data differs from the mean. 

\definecolor{grayLow}{rgb} {0.5,0.5,0.5}
\definecolor{grayMedium}{rgb} {0.7,0.7,0.7}
\definecolor{grayHigh}{rgb} {0.9,0.9,0.9}

\begin{table*}[h!]
\centering
\caption{2022 DORA report performance levels \cite{stateofdevopsreport22}}
\label{tab:performancelevels}
\begin{tabular}{|p{2.5cm}||p{2.5cm}|p{2.5cm}|p{2.5cm}|p{2.5cm}|} \hline
\textbf{Performance Level}   &  \textbf{Deployment Frequency} & \textbf{Lead Time For Changes} & \textbf{Time to Restore Service} & \textbf{Change Failure Rate} \\ \hline \hline
\rowcolor{grayLow} Low & Between once per month and once every 6 months & Between one month and 6 months & Between one week and one month & 46-60\% \\ \hline
\rowcolor{grayMedium} Medium & Between once per week and once per month & Between one week and one month & Between one day and one week & 16-30\% \\ \hline
\rowcolor{grayHigh} High & on-demand (multiple deploys per day) & Between one day and one week & Less than one day & 0-15\%  \\ \hline
\end{tabular}
\end{table*}

\begin{table*}[h!]
\centering
\caption{OSS projecs results and DORA performance levels: Mean / Standard deviation}
\label{tab:performanceresults}
\begin{tabular}{|p{2.0cm}||p{3.0cm}|p{3.0cm}|p{3.2cm}|p{2.3cm}|} \hline
\textbf{Project}   &  \textbf{Release Frequency (mean / standard deviation)} &
\textbf{Lead Time For Released Changes} &
\textbf{Time To Repair Code} & \textbf{Bug Issues Rate} \\ \hline \hline
Angular & \cellcolor{grayMedium}{6.09 days / 3.28 days} & 
 \cellcolor{grayHigh}{3.41 days / 2.36 days} & \cellcolor{grayLow}{79.22 days / 94.11 days} & \cellcolor{grayHigh}{3.84 \% } \\ \hline
Kubernetes & \cellcolor{grayMedium}{5.87 days / 10.75 days}& \cellcolor{grayMedium}{14.31 days / 10.24 days} & \cellcolor{grayLow}{96.24 days / 114.67 days} & \cellcolor{grayLow}{48.45 \%}   \\ \hline
Tensorflow & \cellcolor{grayMedium}{10.91 days / 24.55 days} & \cellcolor{grayLow}{38.72 days / 27.99 days} & \cellcolor{grayLow}{176.09 days / 215.90 days} & \cellcolor{grayLow}{45.10 \%} \\ \hline
VS Code & \cellcolor{grayMedium}{10.93 days / 6.84 days}  & \cellcolor{grayMedium}{9.62 days / 8.33 days} & \cellcolor{grayLow}{38.40 days / 92.71 days} & \cellcolor{grayMedium}{18.96 \%}\\ \hline
\end{tabular}
\end{table*}


Overall, we provided recommendations based on a community perspective of OSS projects. In the following, we report the results obtained from this research. For each project, we show the values obtained for the metrics and we expose conclusions.

\subsubsection{Release Frequency Results}
For a team to increase its performance level, Release Frequency should be getting higher over time (more releases in the same period of time), as shown in the DORA reports \cite{2023-state-oss-report}, where high performers tend to have a higher Deployment Frequency. However, we measured the mean of the days between two consecutive releases, so the lower the value of the metric, the higher Release Frequency. Not only a high Release Frequency is important but also to release periodically (standard deviation in data can influence results). Having a high Release Frequency allows teams to continually receive valuable and actionable feedback and quickly deliver high-quality software.

The results obtained for each of the projects are reported below:

\begin{itemize}
\item \textbf{Angular:} As we can see in Table \ref{tab:performanceresults}, the Angular project has a Release Frequency of 6.09 days, which means that the average time it takes for Angular to make a release since the last one made is less than a week. They have a proper organizational structure and a big team working on the project. In this case, the release frequency is determined by a high number of releases (31 releases in the given period) which allows the team to introduce fewer changes in each release. It also has a low standard deviation, so the releases are relatively regular through time. The use of CI/CD tools like Circle CI and automated tests makes it easier for the team to reduce technical debt.
\item \textbf{Kubernetes:} Unlike other projects, Kubernetes makes releases less often, but in many cases they release multiple versions in the same day, resulting in various releases. This is, in part, possible due to the fact that an automatized bot makes the releases, using it as a tool instead of assigning a team member, given the fact that this team is not as big as the Angular team. They also have a larger standard deviation, which means that their releases are not as regular as in Angular. They make use of GolangCI-lint, a linter for Go that analyzes the code and allows the identification of code smells, as well as automated tests. However, having a more irregular Release Frequency leads to a higher number of changes introduced in some releases.
\item \textbf{TensorFlow:} With a Release Frequency of 10.91 days, TensorFlow has only 8 releases in the measured period. However, the first release made in the period was in September. The TensorFlow team does not really make releases often (some of them are months apart from each other), but in some cases, they make various releases on the same day. This leads to a higher number of changes introduced in some releases. The standard deviation for this metric is the biggest of the four projects, leading to irregular releases and longer periods of time between releases. They use CI/CD tools like Google Cloud services, as well as automated tests that help reduce technical debt. Sometimes bottlenecks in the process can be found as the team is not able to attend to many issues due to a lower number of members in the team. 
\item \textbf{VS Code:} Although the Release Frequency is similar to the TensorFlow one, these projects' approaches to releases are different. VS Code made more releases during the period and in a more even way. This leads to fewer changes introduced in each release. Their standard deviation shows that they do not have a regular Release Frequency, but it is more regular than projects like Kubernetes and Tensorflow. They use CI/CD tools like Azure DevOps services in their process, as well as automated tests, which allows them to make the process faster and reduce technical debt.
\end{itemize}

\subsubsection{Lead Time For Released Changes Results}
A high Lead Time For Released Changes might be an indicator of inefficiencies in team processes. The results obtained for each of the projects are reported below:

\begin{itemize}
\item \textbf{Angular:} This project has a low Lead Time For Released Changes. As mentioned in the Release Frequency section, fewer changes are introduced in each release, compared to other projects. Their standard deviation shows that it takes similar times to introduce changes in the code. 
There are templates for different types of issues and contributors can use them to fill in the information needed. Anyone can submit an issue or change in the code, but only the Angular team members can review these changes. The high Release Frequency allows the team to introduce changes in production fast.
\item \textbf{Kubernetes:} Kubernetes has a Lead Time For Released Changes of 14.31 days. A lot of changes are introduced in some releases as the Released Frequency is irregular. This leads to a higher standard deviation in the data. In many cases, there are multiple open issues at the same time, so it requires more time to review and work on them, which can be a bottleneck in the process. The guidelines on how to contribute to the project cover many topics: opening an issue, triaging an issue, and submitting pull requests, among others. There are also issue templates for different types of issues. Everyone can submit issues and triage them, so it becomes easier for team members to approach those issues.
\item \textbf{TensorFlow:} TensorFlow has a high Lead Time For Released Changes that can be related to the number of releases made. As explained in the Release Frequency section, TensorFlow makes multiple releases on the same day, but it tends to be a long period between those days (this is shown by the standard deviation of the Release Frequency), so the time between the first issues of the period is closed and the release is made tends to be long. This also leads to a higher number of changes introduced in each release. This standard deviation is the highest due to the irregularity of the releases. Guidelines on how to contribute to the project are clear, but not as extensive as in other projects. They do not include a section on how to open an issue (although there are issues templates).
\item \textbf{VS Code:} VS Code has a big team working on it, supported by Microsoft. This allows them to attend to a high number of issues. There is also a bot working on assigning members of the team to the issues, so they can easily manage them, which leads to a better structural organization. Releases are frequent and more periodical than in other projects, which leads to introducing fewer changes in each release and a lower standard deviation (the time it takes to introduce changes in the code is more regular). The guidelines are very extensive and detailed, covering many topics. There are issues templates that make the process of submitting an issue faster and easier. 
\end{itemize}

\subsubsection{Time To Repair Code Results}
Attending to bugs is important to maintain stability in our code. 
This metric measures the time it takes to introduce a change that fixes a bug. In software services, stability is a priority, but the projects studied in this research have a different context as they are products.

The results obtained for each of the projects are reported below:
\begin{itemize}
\item \textbf{Angular:} Angular has a much higher Time To Repair Code than Lead Time For Released Changes. This indicates that it takes more time for them to introduce changes that solve bugs in production than it takes for them to introduce new features. As explained in the Bug Issues Rate section, one of the causes might be the use of the 'bug' label. There are just a few issues labeled as bugs in the period which results in a lower number of data. When filling an issue with the bug template, it is not automatically labeled as a bug. This can lead to results that are not true to reality. As we can see, the Time To Repair Code standard deviation is higher than the standard deviation of the other metrics. This is because we measure how much time it takes to solve a bug from the moment in which the bug issue reporting it is opened. This leads to bug issues that are opened several months before they are solved. However, as already mentioned, Angular has automated tests, guidelines, CI/CD tools, and a big team that help to reduce the Time To Repair Code.
\item \textbf{Kubernetes:} As in the Angular project, Kubernetes also has a high Time To Repair Code. In many cases, a bot informs that the project lacks enough contributors to adequately respond to all issues. It takes months to attend to some issues and others are closed due to inactivity. The standard deviation is high and there is a big variance in data. This can be an indicator of a poor structural organization or a lack of team members.
\item \textbf{TensorFlow:} With the highest Time To Repair Code, TensorFlow takes the longest time to introduce changes that solve bugs in releases out of the four projects of this research. As we explained in the Release Frequency section, TensorFlow made the first release of the measured period in September, which leads to a long time without releasing code (for that same reason they also have the highest Lead Time For Released Changes) and more changes to introduce in a single release. The standard deviation is the highest one and it shows that the mean does not truly represent the actual data, as the variance is really high. Their team members developed a chart (using GitHub projects) in which they organize the changes that need approval, but it seems that there are a lot of changes and not many team members reviewing them, as many change reviews are assigned to the same team members.
\item \textbf{VS Code:} VS Code has the lowest Time To Repair Code. The team members usually attend fast to bug issues. They have the lowest standard deviation as well, which can indicate that most bug issues are attended fast. Automated tests, well-developed guidelines, CI/CD tools, and a high number of members in their team help VS Code to reduce their Time To Repair Code.
\end{itemize}

\subsubsection{Bug Issues Rate Results}
As we defined it, Bug Issues Rate is measured as the percentage of total issues created in the period that are labeled as bugs. A low Bug Issues Rate indicates a low number of bugs in proportion to other types of features, while a high Bug Issues Rate indicates the opposite.

The results obtained for each of the projects are reported below:
\begin{itemize}
\item \textbf{Angular:} Although a low number of bugs can reflect a good development and operations process, this not seems to be the case. There are issues describing bugs that do not have the bug label on them. Although the guidelines on reporting bugs are clear, not every bug reported is classified as one by team members and it makes it more difficult to keep track of the bugs that are being reported. On the other hand, tests and CI/CD tools help reduce the number of bugs and code smells in the project.
\item \textbf{Kubernetes:} Kubernetes has the highest Bug Issues Rate of this research in the given period. All of this bugs have the kind/bug label. This shows that the community follows the guidelines and make good use of the appropiate labels. These guidelines are clear and contributors follow them properly.
\item \textbf{TensorFlow:} TensorFlow has a high Bug Issues which also indicates a good use of the labels provided to classify the issues. As the other projects in this research do, TensorFlow uses many labels to help describe and classify the issues as accurately as possible.
\item \textbf{VS Code:} There are many issues opened and closed every day in the VS Code repository. Although the community used the label to classify issues, it is not needed in many cases, as the project has 450 labels that allow the users to classify the issues more accurately and give a more realistic view of the issues labeled as unknown bugs.
\end{itemize}

\subsection{Summary of the Results}


With our analysis, we confirmed that the benchmarking metrics we propose in this feasibility study can be applied to OSS projects. We show it by applying them to benchmark four popular OSS projects, i.e. Angular, Kubernetes, TensorFlow and VS Code.

We see that Angular and VS Code stand out for having a well-organized structure and sufficient team members, which contributes to effective issue management and resolution. They also adopt a strategy of frequent small updates instead of fewer large updates. VS Code also has many labels for issue classification and uses bots for issue assignment, which is a common practice among all projects.

All projects use bots in some form, such as Angular and TensorFlow using bots for code committing, while Kubernetes is unique in having a bot specifically for code merging and test running. The other projects use workflow automation tools for tasks like integration and code deployment. In terms of issue management, Angular TensorFlow and VS Code use bots for user assignment. TensorFlow also has bots for creating releases.

Angular, TensorFlow and VS Code use GitHub Actions as a workflow automation tool, while each project uses a different tool for code integration: Angular uses Circle CI, TensorFlow uses Google Cloud Services, and VS Code uses Azure Pipelines.

As a result of our analysis, we can conclude that for \textit{Release Frequency} a medium level of performance was observed across the four projects. In terms of \textit{Lead Time For Released Changes}, Angular had a high level of performance, Kubernetes and Visual Studio had a medium level, and TensorFlow had a Low level.

For \textit{Time To Repair Code}, a low level of performance was observed in all four projects. Regarding \textit{Bug Issues Rate}, Angular had a high level of performance (although not reliable due to poor issue tagging), VS Code had a medium level, and Kubernetes and TensorFlow had a low level.



As for the potential influence of DevOps practices on OSS project performance, it seems plausible that the widespread use of bots and workflow automation tools, such as Kubernetes' distinctive bot for code merging and test execution, might play a role in the observed medium Release Frequency performance across all four projects.

It is possible that Angular and VS Code's well-structured organization, adequate team size, and regular minor updates could be contributing factors to their high and medium Lead Time For Released Changes performances, respectively. The utilization of labels and bots for issue management in VS Code might also be a factor in its medium performance for this metric.

The performance of stability metrics (Time To Repair Code and Bug Issues Rate), was generally low across the four projects. This can be due to the lack of an operations service in OSS projects and low community pressure as a result. However, Visual Studio's use of labels and bots for issue management helped to achieve a medium performance in \textit{Bug Issues Rate}. Meanwhile, incomplete bug classification in Angular made tracking bugs challenging.

DevOps practices like automation tools or effective management can improve OSS project performance in terms of throughput metrics. However, the nature of OSS projects makes challenging to them to achieve high levels of stability.


\subsection{Threats to Validity}
We followed the proposal by Wohlin et al. \cite{wohlin2012experimentation} to describe the threats to validity that might bias our performance benchmarking study:

\begin{itemize}
\item \textbf{Conclusion Validity Threats:}
Regarding the reproducibility of the results, all the steps followed to carry out this performance evaluation of OSS projects were detailed in order to ensure their reproducibility. Furthermore, in order to minimize the subjectivity of the actions taken, these actions were checked independently by each of the authors of this work. In addition, regarding the performance results obtained, software measurement instruments were used, therefore, the results obtained are more objective than the surveys carried out in the previous State of DevOps Reports (lines of code are more reliable than surveys since they do not involve human judgment).The material used to make the measurements shown in this study are available, so the results can be replicated.

\item \textbf{Construct Validity Threats:}
The practices and tools to be checked in repositories were defined with reference to reports that have been published in recent years by Puppet Labs and DORA. All authors reviewed this information in order to provide a better understanding of the extent to which DevOps practices affect the performance of OSS projects.

Concerning the validity of the selected sample of projects, in order to ensure that they were sufficiently significant, we defined a number of features regarding their size (number of contributors, number of issues, number of pull requests) and lifetime (at least 8 years of activity).

One of the main problems of this study was the difficulty in applying the metrics due to the  workflow practices variability applied in the different repository projects. This was mitigated by adjusting the definition of each of the four metrics, taking care that each metric respected the scope of the benchmark definitions but was tight enough to be calculated in each case.

\item \textbf{Internal Validity Threats:}
As regards the effect caused by a bad design of the metrics and artifacts used for the study. It should be noted that all adjustments made were within the scope of the definitions of the four metrics. These adjustments were implemented in the measuring instrument software used. The adjustments were checked by all authors to ensure that the results obtained were consistent.

\item \textbf{External Validity Threats:} 
Recent State of DevOps reports recognizes the importance of context in the performance evaluation of organizations. The results obtained might not be fully generalizable because the projects evaluated have different purposes and objectives, and this might affect the context, and consequently, the approach to performance evaluation. In addition, all analyzed OSS projects are on GitHub platform repositories. Therefore, we recognize that our results are limited to the context of these projects. To generalize the results, it is necessary to carry out more studies as future work, both replication, and evaluation of other and different nature projects. Nevertheless, we believe that this study provides an adequate basis and coverage of the issue addressed, as the projects evaluated have the same features with respect to their size and lifetime.
\end{itemize}

\section{Gaps and Challenges}
During our work, we identified a series of challenges that can be addressed in order to expand the knowledge obtained in this study. These challenges pertain to gaps such as (G1) the lack of clarity in defining concepts, (G2) the need to establish and validate benchmark metrics, and (G3) the lack of standards in DevOps practices in the context of OSS projects.

\subsection{G1 - Lack of clarity in defining concepts}

\begin{itemize}

\item \textbf{Challenge 1. Creation of common vocabulary.} The lack of clarity in the vocabulary used to mention concepts can lead to imprecise results. This also makes the communication difficult, which can lead to confusion when transmitting ideas and information. In this study we define the concepts of release, deployment and deliver, which we not find clear in the DORA reports. When talking about DevOps practices and processes, the concepts should be clear so they can be applied correctly and the performance of the projects can be measured avoiding confusion.

\end{itemize}

\subsection{G2 - Need to validate benchmark metrics}

\begin{itemize}

\item \textbf{Challenge 2. Validation of benchmark metrics and exploration of new ones.} The metrics proposed in this study were applied to four different OSS projects, but they have not been validated yet by experts. These metrics should be validated to assure they can be applied to obtain the desired results. It is also important to explore different options. The development of new metrics can lead to more accurate results that are more relevant for certain contexts.

\item \textbf{Challenge 3. Research on the need of new benchmark categories for OSS projects.} In this study, we proposed the use of the DORA performance levels to benchmark the results obtained after measuring the performance of the four OSS projects. This decision was made on the assumption that DORA organizations rely in many cases in OSS projects, so their performance can be classified using the same categories. However, it is important to measure and determine the relationship and the dependency of DORA organizations on these projects. A new set of categories might be needed to classify OSS projects.

\end{itemize}

\subsection{G3 - Lack of standards in the context of DevOps practices}

\begin{itemize}

\item \textbf{Challenge 4. Standarize DevOps practices for OSS projects.} We encountered the problem of measuring different projects as each project carries their processes on their own way. This makes it more difficult to develop automated tools to measure the performance of these projects, as the data needed to measure the metrics is not always available in the same way. Developing standard for this projects does not only solve this issue, but it also helps contributors to understand the way in which different projects are developed, making it easier to contribute to different OSS projects.

\item \textbf{Challenge 5. Study the impact of DevOps practices in OSS projects performance.} In order for the developed benchmark and metrics to be helpful, it is important to study the impact of DevOps practices in OSS projects and their performance. It is important to know which practices affect the performance and in which way they do, so improving performance actions can be developed. 

\end{itemize}




\section{Related work}
We are interested in research studies that show the relationship between DevOps practices and improved performance in OSS projects. Additionally, we value research that offers benchmarking methods (such as the State of DevOps Reports) to compare the performance of different projects, teams, or organizations. These benchmarks can help us understand how different practices and tools impact performance and identify the best ways to optimize performance in OSS projects. In addition, in order to better understand the extent to which DORA reports fit the nature of OSS projects, we are interested in identifying research results that find in the properties of OSS projects any differentiation in terms of the impact that DevOps practices might have on their performance. 

\subsection{Evaluating Performance in OSS Projects}
Different studies can be found in the literature on the performance evaluation of OSS projects. Some of the common features that we can find in the literature describing OSS projects are \cite{mockus2002two}:
\begin{itemize}
\item \textbf{Large number of free contributors:} OSS systems are built by potentially large numbers (i.e., hundreds or even thousands) of volunteers.
\item \textbf{Geographically distributed development}: Developers work in arbitrary locations, rarely or never meet face-to-face, and coordinate their activity with the support of online development management tools.
\item \textbf{In some cases, supported by companies:} 
There are a number of OSS projects that are supported by companies and some participants are not volunteers. 
\item \textbf{Distributed design-build works:} There is no explicit system-level design, or even detailed design.
\item \textbf{Project planning and tasks assigment:} Developers undertake the work they choose to undertake. There is no project plan, schedule, or list of deliverables. In addition, work is not assigned. Instead, there is a backlog of new features and bug issues that represent users demands from which voluntary contributors choose the work they undertake. They fix and propose code changes on these new features or issues.
\end{itemize}

Most research on performance in OSS projects focuses on the effect of certain practices and team factors (e.g. productivity, size, dispersion, and experience) on project efficiency and quality \cite{7230299, 6382850, 5676259}. However, none of these studies introduce benchmarking methods, making it difficult to compare improvement values across OSS projects using similar practices and tools. In highly variable contexts like OSS projects, the use of benchmarking methods is crucial. Moreover, different OSS project characteristics can also impact performance. Global Software Development (GSD) and Distributed Software Development (DSD) are important characteristics of OSS projects, as they involve large numbers of contributors working in different locations and coordinating their activity through online development management tools. Empirical studies suggest that technical practices may influence performance.

\begin{itemize}
\item Coordination mechanisms, such as GitHub workflow, seem to be more influential for geographically distributed development (i.e. OSS projects) than for co-located development \cite{10.1145/302405.302455}. In addition, empirical evidence showed that geographical dispersion impacts negatively and temporal dispersion has a mixed effect on team performance \cite{10.1145/2372251.2372274}.  
\item A large number of contributors looking for bugs and errors in OSS projects can find and fix them faster (also known as "Linus' Law") \cite{mockus2002two}.
\item  As contributors works only on the OSS projects they are passionate about, it seems that their creativity is higher and they tend to care more about the code they develop \cite{HERTEL20031159}. Emotions and moods of contributors are important aspects of work experience that influence performance in OSS projects \cite{weiss1996affective}. In addition, the nature of OSS projects and communities makes it possible to find and contact project leaders much more easily than with traditional software vendors \cite{4273072}. 
\end{itemize}

Empirical evidence supports the use of practices and tools to enhance the efficiency and effectiveness of OSS projects. However, results vary based on the specific characteristics of each project and technology continues to advance. A notable study from 20 years ago, Mockus et al. \cite{mockus2002two}, analyzed data from Apache web server and Mozilla browser and discussed the potential for high-performing OSS and commercial hybrids. However, due to the passing of time, the practices and tools discussed in the study may not be considered high-performance for OSS projects today.

A more current work is the work of Santos et al.\cite{Santos_2022} which collected CI sub-practices (Build Duration, Build Activity, Build Health, Time to Fix a Broken Build, Commit Activity) of 90 relevant OSS projects for a period of 2 years and perform a qualitative analysis to understand whether CI best practices are related to a higher quality of projects. Joy et al. \cite{8479462} investigated the influence of OSS project characteristics on their performance using data from the GitHub repository in 2017. The study confirmed the importance of number of contributors, project age, project size, and number of languages on performance. The study developed a conceptual model with seven hypotheses and found that characteristics of the project number of languages, project size, number of contributors, and project age have a significant impact on project performance. Another study from these authors \cite{thangavelu2018determinants}, presented a conceptual model of the factors of projects, users, contributors, and property rights that influence the OSS project performance. However, all of these studies provide insight into what influences the efficiency and effectiveness of projects, but none of them allow OSS communities to determine what level of performance they are at. Int these lines, none of these studies provide a benchmarking or classification model to compare projects with each other, beyond identifying differences in practices and applied tools.

Regarding the effects of teams on productivity in OSS projects, Pietrantonio et al. in their study \cite{9229495} proposed the use of open-source game projects to evaluate team performance. They analyzed various aspects such as team composition, working history, workflow, communication, performance, task type, and environment, and found that different games have unique strengths in these areas. In adition, Scott et al. in their study \cite{9226278} investigated the productivity of seven agile software development teams in OSS projects to identify key factors. They utilized velocity and focus factor as proxies to evaluate productivity and analyzed the correlation between productivity and OSS-related factors such as turnover of team members, team stability, and iteration length. Their findings indicated that high team stability and low turnover rates are linked to high-velocity iterations. However, none of these studies offer a benchmark to compare the level of productivity or relative performance between teams and projects.

In addition, research was done to predict the viability and success of OSS projects, which is related to the concept of OSS project performance \cite{singh2010small}. For example, Raja et al. \cite{6127835} found that the viability of a project can predict its survival, and identified factors like programming language, target audience, and project type that affect viability. They also created three dimensions (vigor, resilience, and organization) to evaluate a project's overall viability. Schach et al. \cite{schach} analyzed 122,205 projects from SourceForge to determine success criteria for OSS projects and found that only 25\% of projects with one or more downloads can be considered successful. Bayati \cite{bayati2018understanding} focused on the success of newcomers in OSS communities by analyzing their reputation in social coding environments like GitHub. By using mining software repository techniques, Bayati found the main attributes of successful projects for newcomers. While these studies provide insight into what influences the success of projects, they don't allow OSS communities to determine their own level of performance. There's no way to compare projects to each other and determine how they stack up, just differences in the methods and tools used.

Regarding the empirical evidence about the benefits of the application of DevOps practices in organizations, we can find not only survey studies in the literature but also interviews and case studies to learn more about the benefits and challenges of DevOps \cite{erich2017qualitative, caprarelli2020fallacies,diel2016communication}. However, Erich et al. \cite{https://doi.org/10.1002/smr.1885} identified some gaps of evidence about the effectiveness of principles and DevOps practices and metrics to measure their effectiveness. More specifically, the authors could not find literature in which an organization states that they implemented DevOps while providing quantitative data on observed benefits.

Recently, the practice of benchmarking was applied to compare the performance of OSS projects. The focus is limited to the comparison of performance using bots versus not using bots. Markusse et al. \cite{markusseetal2022} propose using bots as a tool for continuous performance benchmarking. Their research indicates that while tools like GitHub Actions are making bots more available to developers, they are still not widely adopted in OSS projects. Thus, this study confirmed the current gap in terms of the automatic use of tools or bots to measure performance in OSS projects.

In summary, the studies in the literature on evaluating the performance of OSS projects suggest that certain practices such as continuous integration and a high number of contributors can have a positive impact on productivity and quality. However, more research is needed to fully understand the extent to which DevOps practices can optimize performance in OSS projects and to provide a benchmarking or classification model to compare the performance of different projects. State of DevOps Reports provide valuable insights on metrics and benchmarking, but further research is needed to adapt these metrics to the specific nature of OSS projects. Overall, the literature suggests that DevOps practices in OSS projects can lead to improved performance, but more research is needed to fully understand the best ways to optimize performance in these projects. 

\subsection{State of DevOps Reports} 
The State of DevOps Reports are a series of annual reports published by Puppet Labs and Google/DORA on the state of the DevOps movement in the tech industry\cite{kim2021devops, forsgren2018accelerate}. The first report was published in 2011 by Puppet Labs and included input from over 4,000 IT professionals. In 2014, Puppet Labs and DORA began collaborating on the reports, and the number of respondents increased to over 9,200 people from 110 countries. The reports included measures of IT performance, such as Throughput Metrics (Deployment Frequency, Lead Time for Changes) and Stability Metrics (Mean Time to Recover, Change Fall Rate), and also introduced the concept of benchmarking to allow teams to compare their own performance to that of high, medium, and low performers. This collaboration between Puppet Labs and DORA ended in 2018.

In 2018, the Accelerate State of DevOps Reports published by DORA introduced the concept of "elite" performers and added a fifth metric to measure operational performance. In contrast, the 2018 State of DevOps Reports published by Puppet Labs analyzed how organizations and teams evolved to achieve the highest levels of DevOps evolution. 
The model categorized respondents into three evolution categories: "Low", "Medium", and "High". The report also concluded that organizations at the highest levels of DevOps evolution also had fully integrated security practices. In 2021, the report published by DORA included operational practices on site reliability (SRE) and showed that teams that prioritize both delivery and operational excellence have the highest organizational performance. 

In recent years, the State of DevOps Reports also focused on the importance of security practices in improving organizational performance. According to the report, it seems that the biggest predictor of an organization's adoption of development security practices is cultural rather than technical \cite{stateofdevopsreport22}. To address this, DORA considers the SLSA (Supply chain Levels for Software Artifacts) framework \cite{slsaframework}, which provides a common language for describing a series of software supply chain integrity practices organized by levels, and the SSDF (Secure Software Development Framework)\cite{ssdfframework}, which provides a common language for describing secure software development practices to help an organization align and prioritize its secure software development activities with its business/mission requirements, risk tolerances, and resources. The SLSA framework is based on industry-recognized best practices and consists of four levels of increasing assurance, while the SSDF is based on documents from organizations such as BSA\footnote{https://www.bsa.org/}, OWASP\footnote{https://owasp.org/}, and SAFECode\footnote{https://safecode.org/}.

\section{Conclusion and future work}
We proposed significant contributions in the field of DevOps practices and their impact on OSS projects. We demonstrated that the metrics defined by DORA are not fully suitable for OSS projects as their characteristics are different from the ones of the DORA organizations. By clarifying the terminology used in the benchmark and focusing on the continuous release of code and its impact on third parties, rather than continuous deployment and delivery to production, we were able to develop a new benchmarking approach that measures the performance of OSS projects. The clarification of the terminology includes the differentiation of environments where the metrics are measured, distinguishing three different environments: Local, Development and Operations. Then, this distinction ensures that we measure the performance levels with the right metrics. The definition of the metric depends on the objective of the environment as each one has its own characteristics and context. The development environment is where software is deployed and tested, while the operations environment is where software is delivered to end users as a service. Therefore, the practices applicable in each environment are different, and the description of the metric must be adjusted accordingly. To avoid confusion, we defined the terms "deploy," "release," and "deliver" more precisely.


In addition, we proposed new metrics that are specifically tailored to the issues related to OSS development. These metrics were designed by adapting the DORA metrics to the particular characteristics of OSS projects, providing a more accurate representation of their performance. We propose using Release Frequency and Lead Time For Released Changes to measure throughput in OSS projects, and Time To Repair Code and Bug Issues Rate to assess its stability. For each of these metrics, we offered a definition and the formula we applied to calculate their values. We found it important to make clear what formulas are being used and what data is being included in each measurement. In the DORA reports, the definition of the data and concepts involved in each metric is not clear enough and can lead to some confusion, which could eventually lead to imprecise results and conclusions.

We found that, in addition to calculating the mean of the metrics, it is necessary to calculate the standard deviation in order to get more detailed information about the performance, as this can provide valuable insights into the stability of the metrics over time. For example, a high standard deviation in Release Frequency might indicate stability issues or lack of proper planning, while a low standard deviation might indicate good planning and stability. Thanks to the standard deviation, we found that the periodicity of releases have a great impact on some metrics such as Lead Time For Released Changes and Time To Repair Code. When the standard deviation of the Release Frequency is lower (more regular releases), the values obtained for the Lead Time For Released Changes and Time To Repair Code metrics are better, showing a better performance.

After measuring the metrics for the four selected OSS projects (Angular, Kubernetes, TensorFlow and Visual Studio Code), we obtained interesting results and information. The Release Frequency for these projects was similar, with no big differences. However, the standard deviation for these projects showed great differences. TensorFlow had a much greater standard deviation than the other projects. The Lead Time For Released Changes metric showed interesting results as there are big differences between projects. We found that the projects with a lower standard deviation for the Release Frequency metric, had better results in the Lead Time For Released Changes metric. The Time To Repair Code metric showed really low performance for the four projects, which can be related with the context of OSS projects and their characteristics, as they do not necessarily prioritize the correction of bugs. The Bug Issues Rate metric also showed big differences between projects. We concluded that the categorization and identification of bug issues in the repositories had a great impact on this metric. The Bug Issues Rate metric highly relies on how well bugs are identified, so a better value on this metric does not always represent a better performance.


Our benchmark is based on automatic measurement using project repositories, which reduces the risk of survey-based data collection. We also developed a new tool called Performance-Tracker, which connects to OSS project repositories and applies our performance benchmark. This tool can help practitioners assess their own OSS projects and identify areas for improvement. However, the main challenge we encountered was applying the metrics due to the variability of workflow practices in the different GitHub project repositories. To overcome this challenge, we need to establish best practices that are shared by the community or develop tools that can adapt to different configurations.

Overall, our research has contributed to a better understanding of the performance of OSS projects. We hope that our findings will be useful for future DORA reports and for practitioners looking to improve the performance of their OSS projects. 

To advance the understanding of DevOps practices in the context of OSS projects, it is crucial to address the challenges identified in this paper. Collaboration and cooperation among stakeholders in the OSS community will be essential to overcome these challenges.

One of the main challenges is the lack of clarity in defining key concepts. To address this, future work will focus on developing a common vocabulary for DevOps concepts and validating it with experts in the field. This will improve communication and reduce confusion among stakeholders, leading to more precise results.

Another challenge is the need for reliable metrics and benchmarks that accurately reflect the performance of OSS projects. Future work will focus on validating metrics and exploring new ones that are more relevant or applicable in different contexts. Developing standardized DevOps practices for OSS projects will also be essential to improve performance measurement and make it easier for contributors to participate in different projects.

Furthermore, it is important to research the need for new benchmark categories that more accurately classify and measure the performance of OSS projects. This will involve exploring the relationship between OSS projects and DORA organizations and determining whether existing benchmark categories are relevant and applicable to a wide range of OSS projects.

\section*{Materials}

For the sake of open science, all the material including he tool, the deployment environment and all the rough data  used in this paper is available online at https://github.com/diverso-lab/performance-tracker .

\ifCLASSOPTIONcompsoc
  \section*{Acknowledgment}
\else
  \section*{Acknowledgment} 
\fi

This research is partially funded by the Spanish Ministry of Science, Innovation and Universities under the NICO (PID2019-105455GB-C31), the SUDOQU (PID2021-126436OB-C21,  10.13039/ 501100011033, FEDER, UE), the METAMORFOSIS (FEDER\_US-1381375), the Data-PL projects and the grants for requalification of spanish university system 2021-2023 funded by the European Union - NextGenerationEU. Also with the support of the Generalitat de Catalunya under Grup de Recerca Consolidat IMP, 2021-SGR-01252 and inLab FIB at the UPC.

\ifCLASSOPTIONcaptionsoff
  \newpage
\fi



%


\bibliographystyle{IEEEtran}
\bibliography{references}



%

\begin{IEEEbiography}
[{\includegraphics[width=1in,height=1.25in,clip,keepaspectratio]{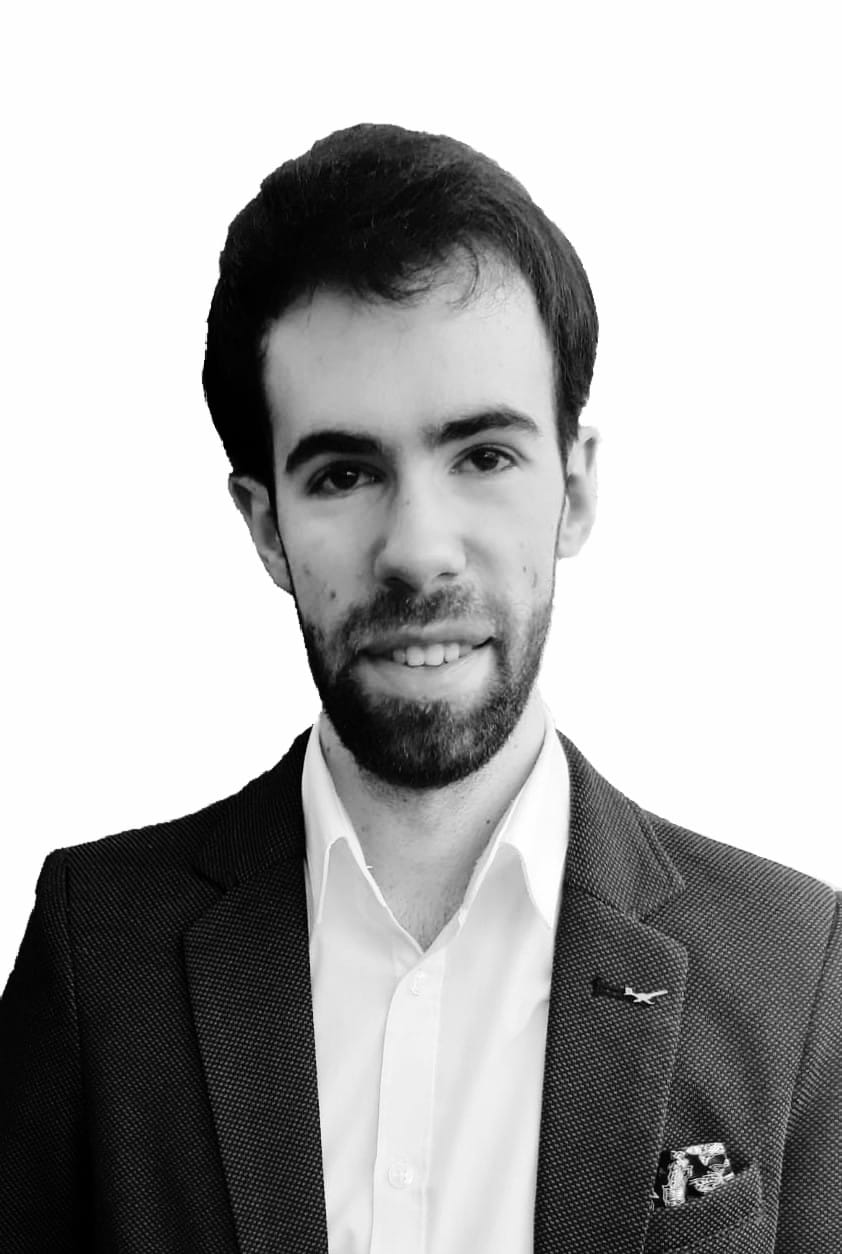}}]
{José Manuel Sánchez Ruiz}
received a degree in computer software engineering from the University of Seville in July 2021. He is currently a researcher with the Deparment of Computing Languages and Systems, University of Seville. He collaborates in studies and researches related to software engineer and software team development, as well as the development of tools, systems and applications.
\end{IEEEbiography}
\begin{IEEEbiography}[{\includegraphics[width=1in,height=1.25in,clip,keepaspectratio]{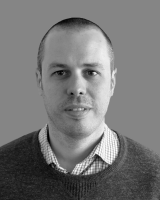}}]{Francisco José Domínguez Mayo}
received a PhD degree in computer science from the University of Seville in July 2013. He is currently an associate professor with the Department of Computing Languages and Systems, University of Seville. He collaborates with public and private companies in software development quality and quality assurance. The focus of his interesting research is on the areas of continuous quality improvement and quality assurance on software products, and software development processes.
\end{IEEEbiography}
\begin{IEEEbiography}
[{\includegraphics[width=1in,height=1.25in,clip,keepaspectratio]{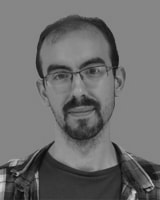}}]
{Xavier Oriol} received a PhD degree in computer science in 2017 from Universitat Politècnica de Catalunya (UPC) with International mention (Cum Laude). He is a researcher and project leader at inLab FIB, and teaches software engineering courses in the same faculty. His research interests include incremental integrity checking, semantics, and automated reasoning on conceptual schemas.  
\end{IEEEbiography}
\begin{IEEEbiography}
[{\includegraphics[width=1in,height=1.25in,clip,keepaspectratio]{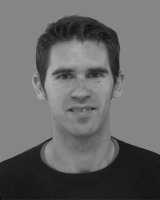}}]
{José Francisco Crespo} is a computer scientist from the Barcelona School of Informatics at Universitat Politècnica Catalunya (UPC). He received his MSc in Computer Architecture, Networks, and Systems from UPC. He has experience in R+D projects and developing agile software projects. He is interested in applying technology to collaboration environments and mobility. Currently, he is involved in the development of web projects and mobile apps
\end{IEEEbiography}
\begin{IEEEbiography}
[{\includegraphics[width=1in,height=1.25in,clip,keepaspectratio]{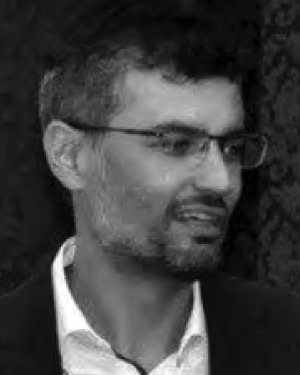}}]
{David Benavides} received the B.S. degree in information systems from the Institute Superieur d’Electronique de Paris, France, in 2000, and the M.Sc. degree in computer engineering and the Ph.D. degree in software engineering from the University of Seville, Spain, in 2001 and 2007, respectively. He is currently a Full Professor with the University of Seville. His main research interests include software product line and artificial intelligence applied to engineering education.
\end{IEEEbiography}
\begin{IEEEbiography}
[{\includegraphics[width=1in,height=1.25in,clip,keepaspectratio]{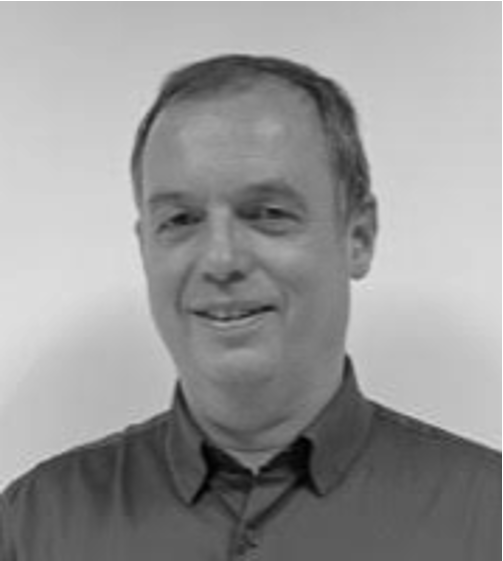}}]
{Ernest Teniente}  is full professor in Software Engineering at the Universitat Politècnica de Catalunya (UPC). He is also Director of inLab FIB and the head of the Information Modeling and Processing (IMP) research group, both at the UPC. His research interests include ontologies and conceptual modeling, business process modeling, semantics and automated reasoning, automatic code generation, integrity constraints enforcement, and data integration.
\end{IEEEbiography}







\end{document}